\documentclass[10pt]{iopart}

\usepackage{iopams}
\usepackage{amssymb}
\usepackage{braket}

\usepackage{comment}
\usepackage[T1]{fontenc}
\usepackage{amsfonts}
\usepackage{graphicx}
\usepackage{pgf} 
\usepackage{dcolumn}
\usepackage{bm}
\usepackage{bbold}
\usepackage{soul}
\usepackage{cancel}
\usepackage{latexsym}
\usepackage{color}
\usepackage{xcolor}
\usepackage{dsfont}
\usepackage{enumerate}
\usepackage{ulem}
\usepackage{verbatim}
\usepackage[separate-uncertainty=true]{siunitx}
\usepackage[english]{babel}
\usepackage{hyperref}
\hypersetup{
citecolor=blue,
colorlinks=true,
urlcolor=magenta,
}

\newcommand{\bb}[1]{\left( #1 \right)}
\newcommand{\fluct}[1]{\delta^2 #1 }

\newcommand{\gmne}{\Gamma\bb{E,\,N}}
\newcommand{\gmneex}{\Gamma_{\rm ex}\bb{E,\,N_{\rm ex}}}

\newcommand{\nex}{N_{\rm ex}}
\newcommand{\kB}{k_{\rm B}}
\newcommand{\oppsi}{\hat{\Psi}\left(\bm{r}\right)}

\newcommand{\meanv}[1]{\langle {#1} \rangle }
\newcommand{\opr}[1]{\widehat{#1}}
\newcommand{\dagopr}[1]{\widehat{#1}^{\dagger}}
\newcommand{\bo}[1]{{\mathbf{\bm{#1}}}}
\newcommand{\mc}[1]{{\mathcal{#1}}}

\newcommand{\wb}[1]{{\overline{#1}}}
\newcommand{\ve}{\varepsilon}
\newcommand{\vej}{\varepsilon_j}
\newcommand{\nonu}{\nonumber}
\newcommand{\eqn}[1]{(\ref{#1})}
\newcommand{\eq}[2]{\begin{equation}\label{#1}{#2}\end{equation}}

\newcommand{\ddj}{d (\epsilon_j)}

\definecolor{darkbrown}{rgb}{0.6, 0.2, 0.15}

\newcommand{\eqref}[1]{(\ref{#1})}
\newcommand{\binom}[2]{\left(\begin{array}{c} #1 \\ #2 \end{array} \right)}
\newcommand{\tfrac}[2]{\frac{#1}{#2}}

\begin{document}

\title{
On the fluctuations of the number of atoms in the condensate
}

\author{Maciej B. Kruk$^{1,3}$, Piotr Kulik$^{3}$, Malthe F. Andersen$^2$, 
Piotr Deuar$^{1}$, Mariusz Gajda$^{1}$, Krzysztof Paw\l{}owski$^3$\footnote{Author to whom correspondence should be addressed.}, Emilia Witkowska$^1$, Jan J. Arlt$^2$, Kazimierz Rz\k{a}\.zewski$^3$
}
\address{$^1$ Institute of Physics PAS, Aleja Lotnikow 32/46, 02-668 Warszawa, Poland}
\address{$^2$ Center for Complex Quantum Systems, Department of Physics and Astronomy, Aarhus University, Ny Munkegade 120, DK-8000 Aarhus C, Denmark.}
\address{$^3$ Center for Theoretical Physics, Polish Academy of Sciences, Al. Lotnik\'{o}w 32/46, 02-668 Warsaw, Poland.}

\ead{pawlowski@cft.edu.pl}
\vspace{10pt}
\begin{indented}
\item[]February 2025
\end{indented}

\date{\today}

\begin{abstract}
Bose-Einstein condensation represents a remarkable phase transition, characterized by the formation of a single quantum subsystem. As a result, the statistical properties of the condensate are highly unique. In the case of a Bose gas, while the mean number of condensed atoms is independent of the choice of statistical ensemble, the microcanonical, canonical, or grand canonical variances differ significantly among these ensembles.
In this paper, we review the progress made over the past 30 years in studying the statistical fluctuations of Bose-Einstein condensates. Focusing primarily on the ideal Bose gas, we emphasize the inequivalence of the Gibbs statistical ensembles and examine various approaches to this problem. These approaches include explicit analytic results for primarily one-dimensional systems, methods based on recurrence relations, asymptotic results for large numbers of particles, techniques derived from laser theory, and methods involving the construction of statistical ensembles via stochastic processes, such as the Metropolis algorithm.
We also discuss the less thoroughly resolved problem of the statistical behavior of weakly interacting Bose gases. In particular, we elaborate on our stochastic approach, known as the hybrid sampling method.
The experimental aspect of this field has gained renewed interest, especially following groundbreaking recent measurements of condensate fluctuations. These advancements were enabled by unprecedented control over the total number of atoms in each experimental realization. Additionally, we discuss the fluctuations in photonic condensates as an illustrative example of grand canonical fluctuations.
Finally, we briefly consider the future directions for research in the field of condensate statistics.
\end{abstract}

\submitto{\RPP}
\ioptwocol

\tableofcontents

\newpage

\section{Introduction}
Fluctuations are an important property of quantum systems in general. In many cases they are well understood, however, the complete description of the fluctuations of interacting Bose gases remains an outstanding challenge. Since the field of ultracold gases is rapidly maturing, such systems are now at a state, where quantum-based sensor technology is approaching applications outside of laboratory environments, and the understanding of the fluctuations is ever more important.

In particular, Bose-Einstein condensates (BECs) have become a pivotal tool in advancing quantum simulations~\cite{Bloch2012}, and one might expect that a thorough understanding of their properties has been reached. Yet, despite significant theoretical efforts, the atom number fluctuations between the thermal and condensed components in an interacting Bose gas at relevant densities remain elusive. Ideally, such quantum systems can be characterized by all moments of their probability distribution, which have not yet been fully obtained for large interacting BECs~\cite{Politzer1996,Navez1997,Giorgini1998,Meier1999,Kocharovsky2006}.

\begin{figure}[h]
    \centering
    \includegraphics[width=\linewidth]{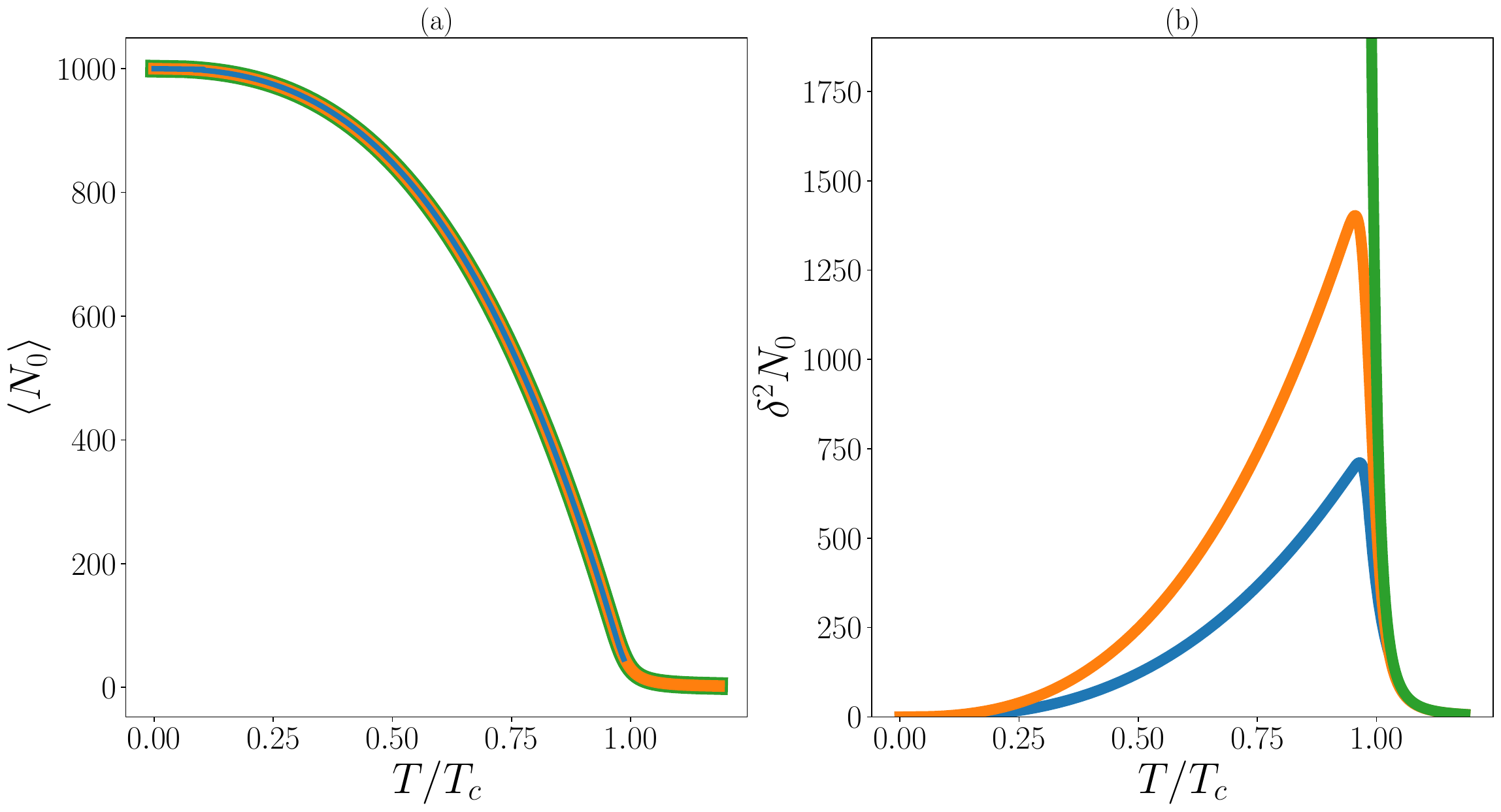}
    \caption{
    Illustration of the average number of condensed atoms (left) and its variance $\fluct{N_0}$ (right) as a function of temperature $T$ in the three archetypical statistical ensembles. The choice of ensemble is irrelevant for $\langle N_0\rangle$. On the other hand, when the temperature of an ultracold Bose gas is lowered towards the critical temperature $T_c$, a grand canonical ensemble (green) predicts unphysically large fluctuations. A canonical ensemble (orange) does not suffer from this problem and has fluctuations peaking just below $T_c$. A microcanonical ensemble shows the same temperature-dependent trend but with quantitatively lower fluctuations. Data for $N=1000$ non-interacting atoms in an isotropic harmonic trap. The right panel is adopted from \cite{Kruk23} ( \href{https://creativecommons.org/licenses/by/4.0/}{CC BY 4.0}).
    \label{fig:graphical-abstract}}
\end{figure}
These fluctuations present a complex problem rich with intriguing challenges~\cite{Kocharovsky2006,Idziaszek2005,Christensen2021,Kocharovsky16,Touchette15,Yukalov24}. In the forties of the last century, E.~Schr\"odinger~\cite{Schroedinger1989} was probably the first to note that the grand canonical ensemble implies unphysically large fluctuations of the condensate population below the critical temperature~\cite{Ziff1977}, a phenomenon known as the grand canonical catastrophe~\cite{Grossmann1996,Holthaus1998}. This highlights one of the few cases where different statistical ensembles yield vastly different results, necessitating the use of canonical or microcanonical approaches. See Fig.~\ref{fig:graphical-abstract}, which contrasts the enormous spread of condensate fluctuations between different ensembles with their good agreement for the mean condensate fraction.

Renewed interest in the statistics of the condensate was brought by the experimental discovery of BEC in ultracold atomic gases in 1995. The asymptotic expression for the variance of the condensate population in the noninteracting Bose gas under the canonical ensemble was derived by Politzer~\cite{Politzer1996}. In parallel, numerical techniques were invented for the exact calculations for the finite number of particles~\cite{Weiss1997}. 
While the canonical ensemble assumes an external reservoir exchanging energy with the system, an assumption invalid for isolated ultracold gases, nonetheless it  provides a first intuitive understanding of the situation as shown in Fig.~\ref{fig:graphical-abstract}.

A description of experiments with ultracold atomic gases requires using the microcanonical ensemble, which is numerically challenging and has yet to be fully compared with experimental conditions. In the microcanonical ensemble, the partition function $\Gamma (E, N)$—representing the number of ways to distribute energy $E$ among $N$ atoms—is key. Its computation for a noninteracting gas in a harmonic trap relates to the classical partition problem, extensively studied by mathematicians such as Leibniz and Bernoulli, with significant breakthroughs by Ramanujan and Hardy a century ago~\cite{Hardy1918}. Notably, it was shown that canonical and microcanonical fluctuations coincide for large atom numbers in the noninteracting 1D gas~\cite{Grossmann1997, Wilkens97}.

The 3D case for the microcanonical ensemble was not solved until 1997~\cite{Navez1997}, when a new ensemble, the Maxwell's demon ensemble, was introduced. In this ensemble, the system exchanges particles with a reservoir while conserving energy, applying to the thermal cloud at low temperatures, where the BEC acts as the reservoir. This model provides an asymptotic expression for microcanonical fluctuations below the critical temperature $T_c^0$ of the noninteracting gas. The variance of the condensed atom number $N_0$ is then given by 
\begin{equation}
\label{eq:Microcan} 
\fluct{N_0} = \left(\frac{\zeta(2)}{\zeta(3)} - \frac{3\zeta(3)}{4\zeta(4)}\right)N\left(\frac{T}{T_c^0}\right)^3, 
\end{equation} 
as a function of temperature $T$ and total atom number $N$ where $\zeta(x)$ is the Riemann zeta function. The second term reflects the reduced fluctuations in the microcanonical ensemble compared to the canonical result, with a 61\% reduction.

Of course, the ensemble description is physically meaningful only in the presence or with a history of interactions. Importantly, interactions are expected to alter condensate atom number fluctuations $\fluct{N_0}$. In the homogeneous case, initial results indicated that interactions suppress fluctuations at low temperatures, due to strong atom pair correlations that limit the degrees of freedom~\cite{Giorgini1998, Meier1999, Zwerger2004}. Such interaction effects have been explored using various methods, including the Maxwell's demon ensemble~\cite{Idziaszek1999}, number-conserving quasiparticle techniques~\cite{Kocharovsky2000a,Kocharovsky2000}, and master-equation methods~\cite{Svidzinsky2006, Svidzinsky2010}, but most approaches either do not fully account for microcanonical conditions or are restricted to low temperatures~\cite{Bhattacharyya2016} and small atom numbers.

Finally, under many conditions below $T_c$, the fluctuations are expected to scale \textit{anomalously} with atom number $\fluct{N_0} \propto N^{1+\gamma}$, with $ \gamma \neq 0 $~\cite{Giorgini1998,Meier1999} in stark contrast to most classical systems where the central limit theorem ensures normal scaling of the fluctuations $\fluct{N} \propto N$ due to the existence of a finite microscopic coherence length scale~\cite{Zwerger2004,Yukalov2005a}. 

The experimental characterization of BEC atom number fluctuations has long been a significant challenge, largely hindered by technical fluctuations in BEC experiments. However, this limitation was recently overcome~\cite{Gajdacz2016}, allowing for the experimental progress that led to the first ever observation of the  fluctuations in the atomic BEC \cite{Kristensen2019, Christensen2021},  detailed at the end of this review. Notably, the measured fluctuations were significantly smaller than predicted by the canonical ensemble \cite{Christensen2021}. In parallel, significant theoretical progress has recently been made and the Fock state sampling and related hybrid sampling techniques bring the description of experimental systems into reach.

In this review we first describe in detail the baseline situation of the ideal gas in its many forms in Sec.~\ref{sec:fluct-ideal-gas}, including a historical survey of results since the 1990s. Sec.~\ref{sec:modern-frameworks} then reviews  numerical approaches used to generate ensembles of samples/realizations for analysis of realistic finite-size cases, with particular attention on Monte Carlo approaches based on a Metropolis algorithm. Sec.~\ref{INT} dives into the fascinating and not fully resolved topic of fluctuations in the weakly interacting condensates, along with again a historical survey of results since the 1990s. Sec.~\ref{sec:experiments} outlines the recent remarkable experimental progress which has allowed direct measurement of the global condensate fluctuations. Finally we conclude and summarize the near- and mid-term future outlook of the topic in Sec.~\ref{sec:perspectives}.

\section{Fluctuations in the ideal gas\label{sec:fluct-ideal-gas}}
The ideal gas fluctuation behaviour underlies the whole topic, and across different dimensionalities, ensembles, and geometries demonstrates already many of the essential features and unusual results of condensate fluctuation statistics, as well as the methodological difficulties involved. Therefore in the first half of the review its features and the methods that can be used to study it will be described in detail, concentrating on the results 
discovered since the 1990s till today.

\subsection{Archetype statistical ensembles}

Statistical mechanics provides a formalism allowing for the description of systems of large numbers of particles in the language of probability theory. It is based on the concept of statistical ensembles, i.e. a set of all possible states of the system subject to external constraints imposed on the system. Every ensemble is characterized by the corresponding partition functions, that determine statistical properties of the gas. 
To quantify the statistics we use throughout the review the probability distribution of the number of atoms outside condensate, denoted with $\mathcal{P} \bb{\nex}$. Of particular interest are the low moments of the distribution 
\begin{eqnarray}
    \meanv{\nex} &=&  \sum_{\nex} \mathcal{P} (\nex) \,\nex \label{eq:average-nex-general}\\
    \meanv{\nex^2} &=&  \sum_{\nex}\mathcal{P} (\nex) \,\nex^2.
    \label{eq:average-nex2-general}
\end{eqnarray}
These statistical moments can be used to compute the average number of atoms in the condensate and the main subject of this review -- the fluctuations:
\begin{eqnarray}
    \left< N_0\right> &=& 
    N-\left< \nex\right> \label{eq:average-n0-general}\\
    \fluct{N_0} &=& \fluct{\nex} = \langle N_0^2\rangle-\langle N_0\rangle^2 
    \label{eq:fluct-n0-general}
\end{eqnarray}
The last equation applies only in the absence of particle exchange with the environment and then expresses the fact that since
the gas comprises only condensed and excited atoms, 
their fluctuations must be equal to each other. It does not hold 
in the grand canonical ensemble as
will be discussed later.

In the formalism of statistical physics, many-body systems at thermal equilibrium can be described by several parameters, which can be divided into extensive ones, such as total energy $E$, number of particles $N$, and volume $V$, and corresponding intensive parameters, such as temperature $T$, chemical potential $\mu$, and pressure $p$.  Ultracold atomic gases, however, are typically trapped in a harmonic potential; they are not uniform, and their volume is not well defined. The role of the volume of the gas container, $V$, is replaced by the external trap frequency $\omega$. In the following, we will focus on equilibrium systems, assuming that the trap frequency (or volume) is fixed. Therefore, we will omit the corresponding variables $\omega$, or $V$ from the formalism presented below, as they do not play a role.

We start by recalling the three archetypical ensembles for describing the ideal Bose gas, the microcanonical, canonical and grand canonical ensembles in Sections \ref{sec:MC}, \ref{sec:CN} and \ref{sec:GC}.
We show the textbook definitions for particular partition functions and complement them with non-standard examples of their analytical evaluation relevant for BECs. The applications of the formalism to the fluctuations of the ideal Bose gas are presented in Section~\ref{section_BEC} and \ref{sec:maxwelldeamon}.

\subsubsection{Microcanonical (MC)}
\label{sec:MC}
If the system is perfectly isolated, it is natural to assume that all  (micro)states of the system consisting of $N$ particles and having total energy $E$ are equally probable. The statistical ensemble assigned to such a system is the microcanonical ensemble (MC). The uniform probability distribution assigned to every state of this ensemble is equal to 
$1/\gmne$ 
where $\gmne$ is the microcanonical partition function, equal to the number of all microstates that meet the constraints of energy and particle number. 
To give a more operational definition we assume that energy is coarse-grained, i.e. the entire range of energies is divided into energy shells $\Delta_E \ll E$ such, that the number of states $N_\Delta$ within a given interval is large $N_\Delta \gg 1$. For large systems, the final results do not depend on the particular choice of~$\Delta_E$. The microcanonical partition function is then:  
\begin{equation}
\Gamma(E,N)=\sum_{\rm{states}}\delta_{\Delta_E}(E-H_N),
\label{micro}
\end{equation}
where $H_N$ is a Hamiltonian of the $N$-particles system, and, according to our discussion on coarse-graining,   $\delta_{\Delta_E}(E-H_N)=1$, if the energy of a given state is in the interval $[E-\Delta_E;E]$ and zero otherwise.

Knowledge of $\gmne$  alone does not suffice to determine the system’s statistical properties directly. Specifically, fluctuations in the atom number within the condensate -- the primary focus of this review -- require an additional quantity $\gmneex$.
This represents the total number of microstates with exactly $\nex$ atoms in excited states at energy $E$. The probability of observing exactly $\nex$ excited bosons is:
\begin{equation}
    \mathcal{P} (E, N_{\rm ex})=\gmneex/\gmne.
    \label{eq:prob-nex}
\end{equation}
This distribution is sufficient to compute the statistical properties of interest following Eqs.~\eqref{eq:average-nex-general}-\eqref{eq:average-n0-general}.

The calculation of the microcanonical partition function $\gmne$ or $\gmneex$ leads inevitably to combinatorial problems, which we illustrate in the conceptually simplest case of an ideal gas trapped in a one dimensional harmonic trap.
In this case, the spectrum of a single particle is, in oscillatory units, $\epsilon_j = j$, \, ($j = 0,\,1\, \ldots$)\footnote{Without loss of generality we are omitting everywhere the shift $\hbar\omega/2$.}.
Therefore each microstate is just a set of $N$ integers, 
being the $N$ ordered values of occupied energy levels,
i.e. $0\leq\epsilon_{j_1} \leq \epsilon_{j_2} \leq \ldots \leq  \epsilon_{j_N}$, and obeying the constraint of fixed energy
\begin{equation}
    \label{eq:paritions-1d-micro}
    E = \epsilon_{j_1} + \epsilon_{j_2} + \ldots + \epsilon_{j_N}.
\end{equation}
The partition function $\gmne$ is the number of the microstates with given energy and number of atoms, so here -- the number of ways of writing the integer $E$ as a sum of $N$ non-negative integers, as in Eq.~\eqref{eq:paritions-1d-micro}.
Although the problem has a simple mathematical presentation via Eq.~\eqref{eq:paritions-1d-micro}, still finding explicitly $\gmne$ is a challenging task, even numerically. It is related to a classic problem in the number theory called the ``partition problem'', studied over centuries by many mathematicians, Leibnitz, Euler, Bernoulli, Hardy, Ramanujan to name a few~\cite{andrews2013}. The partition problem is to find the 
number of partitions  $A(E)$  of an integer $E$
into a sum of \textbf{any number} of \textbf{positive} integers $\epsilon^{(+)}_{j}$
\begin{equation}
 \label{eq:partitions-ramanjuan}
    E = \epsilon_{j_1}^{(+)} + \epsilon^{(+)}_{j_2} + \ldots 
\end{equation}
In the context of our physical problem $\epsilon_{j}^{(+)}$ shall be interpreted as the occupied excited energy levels in one microstate. 
Finally Ramanujan and Hardy found the asymptotic (large $E$) solution $A(E)=\frac{e^{\pi\sqrt{2E/3}}}{4E\sqrt{3}}$\cite{Hardy1918}\footnote{The exact values can be easily evaluated using for instance Mathematica (function {\tt PartitionP[]}). }.

Note that the maximal number of terms on the right-hand side of Eq.~\eqref{eq:partitions-ramanjuan} is $E$ -- this is a partition with all terms equal $1$.  This observation, if we interpret the terms as occupied excited energy levels, means that the number of excited atoms is limited, i.e. $\nex \leq E$. As a consequence, if 
$N\geq E$, then at least $N-E$ atoms, however large $N$ is, must occupy the $0$-energy level, i.e. the ground state. This saturation of the number of excited atoms is reminiscent of the origin of Bose-Einstein condensation, which will be discussed later\footnote{Strictly speaking, there is no condensation in a one dimensional system due to the lack of a critical temperature in the limit $N\to\infty$, but still the number of excited atoms is bounded for any finite fixed energy.}.
 In the case of $N>E$, to every partition of the form Eq.~\eqref{eq:partitions-ramanjuan}, one can add $N_0:=N-N_{\rm ex}$ zeros:
\begin{equation}
E = \underbrace{0 + 0 + \ldots + 0}_{N_0 {\text{ atoms at } \epsilon_0}} + \underbrace{\epsilon_{j_1}^{(+)} +  \ldots \epsilon^{(+)}_{j_{\nex}} }_{\text{$\nex$ excited atoms}}
\end{equation}
to obtain a decomposition of the form of Eq.~\eqref{eq:paritions-1d-micro}. 
This implies that for $N\geq E$, $\gmne$ does not depend on $N$ and is just equal to $A(E)$.

\begin{figure}
    \centering
    \includegraphics[width=\linewidth]{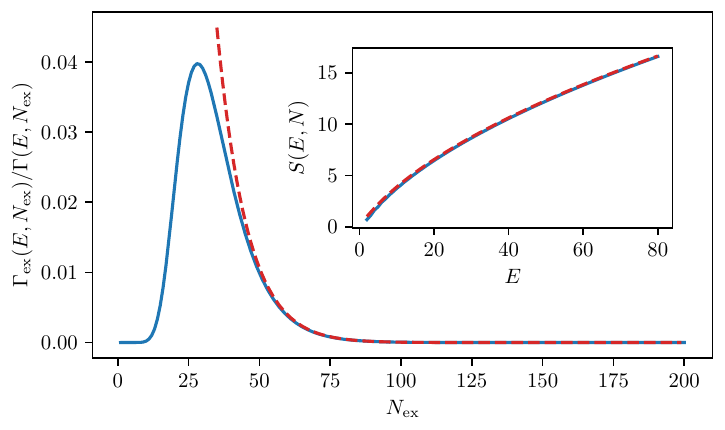}
    \caption{The microcanonical probability distribution of having $\nex$ atoms in a 1d harmonically trapped Bose gas:  
    exact result (blue solid) obtained via Eq.~\eqref{eq:1d-recurrence} and asymptotic formulas for the number of restricted~\cite{Szekeres1953Jan} partitions (dashed red). The inset shows comparison of the entropy $S(E, N):=\kB \ln \gmne$, where $\gmne$ is computed via exact method (blue line) or approximated with the asymptotic formula $A(E)$~\cite{Hardy1918} (red dashed line). Parameters $E=200$ and $N>E$.\label{fig:fig1}}
\end{figure}

Finding the statistics of the number of excited atoms, $\mathcal{P} (E,\nex)$, requires the function $\gmneex$.
In the case of a 1D harmonic oscillator, $\gmneex$ is the number of ways of writing the energy $E$ as a sum of exactly $\nex$ positive integers, 
known as the problem of restricted partitions. There are asymptotic formulas for $\gmneex$ in some specific cases~\cite{szekeres1951, Szekeres1953Jan}, but the general problem remains unsolved~\cite{Bridges2020Dec}.
The $\gmneex$ can, however, be computed numerically using various recurrence relations, among which the simplest -- for the 1D harmonic oscillator -- is: 
\begin{eqnarray}
    \Gamma_{\rm ex}\bb{E, \nex} &=  \Gamma_{\rm ex}\bb{E - 1, \nex-1} +\nonumber \\ 
    &\Gamma_{\rm ex}\bb{E - \nex, \nex}.
\label{eq:1d-recurrence}
\end{eqnarray}
Other recurrence relations for $\gmneex$ in this special case of a 1D gas in a harmonic potential are reported in~\cite{Weiss2002Aug}. Fig.~\ref{fig:fig1} illustrates the accuracy of the asymptotic formulas of~\cite{Szekeres1953Jan} and~\cite{Hardy1918} by comparing them with exact results for entropy and probability distribution obtained with Eq.~\eqref{eq:1d-recurrence}.
Although the asymptotic formulas are sufficient to estimate entropy as a function of energy, and therefore also the temperature, they are not sufficiently accurate to give reliable statistics of $\nex$.
The analogy to number theory, and the results of mathematicians (e.g. Erd\"os~\cite{Erdos1941Jun}), allowed the Authors of~\cite{Grossmann1996} to study the fluctuations of the number of ground state atoms in the 1D harmonic potential analytically and show e.g. that $\Delta N_0/N\approx \frac{\pi}{\sqrt{6}\ln N} \frac{T}{T_0}$, with a characteristic temperature 
$T_0:=N/\ln N$. Further works~\cite{Weiss2002Aug} studied higher moments of the distribution $\mathcal{P} (E,\, \nex)$. 

The above correspondence between the microcanonical partition function and the classic problem in number theory is due to the simple spectrum of the 1D harmonic oscillator. 
In the case of a box, the problem reduces to finding the function $A_2(E)$, which represents the number of ways to partition $E$ as a sum of squares. An asymptotic formula for $A_2(E)$ was first proposed by Hardy in his foundational 1918 paper~\cite{Hardy1918}. This formula was later proved by E. Wright in 1934~\cite{Wright1934Jan} and subsequently simplified by R. Vaughan in 2015~\cite{Vaughan2015Apr}. It is:
\begin{equation}
    A_2(E) \propto \exp\bb{3 \sqrt[3]{E\,\Gamma(3/2)^2 \zeta(3/2)^2/4 } }
\end{equation}
In higher dimensions or for more complex spectra, the problems of finding asymptotic solutions become increasingly difficult and less explored.

On the other hand there exist recurrence relations for $\Gamma_{\rm ex}$ for any energy spectrum. For instance, one can build on the recurrence relation for another object
$\Gamma_{\rm ex} \bb{N, E, \vej^{\rm c}}$,  the number of partitions of energy $E$ into $N$ terms, none of them  exceeding $\vej^{\rm c}$. Here $\vej^{\rm c}$ serves as the energy cut-off, limiting the available spectrum. The recurrence relation over the energy cut-off is~\cite{thesisIdziaszek}:
\begin{eqnarray}
\lefteqn{\Gamma_{\rm ex} \bb{N, E, \vej^{\rm c}}  =} && \label{bigrecurrence}\\
&&\sum_{N_j=0}^N\Gamma_{{\rm ex}} \bb{N-N_j, E - N_j \vej^{\rm c}, \ve_{j-1}^{\rm c}}
 \binom{d(\vej^{\rm c}) + N_j - 1}{N_j - 1}.\nonumber
\end{eqnarray}
The iterator $N_j$ in the sum is the number of particles at the cut-off energy level $\vej^{\rm c}$. Due to the possible degeneracy $d(\vej^{\rm c})$ of this level, the $N_j$ atoms can be distributed in $\binom{d(\vej^{\rm c}) + N_j- 1}{N_j - 1}$ ways. The remaining $N-N_j$ atoms occupy levels not larger than $\ve_{j-1}^{\rm c}$ and carry the energy $E - N_j \vej^{\rm c}$. The recurrence starts with $\Gamma_{\rm ex} \bb{N, E, \ve_{0}} = \delta_{N, \,0} \delta_{E, \,0}$ and ends once $\vej^{\rm c}$ reaches the total energy $E$, because $\Gamma_{\rm ex} \bb{N, E, E} \equiv \Gamma_{\rm ex} \bb{N, E}$.

The recurrence Eq.~\eqref{bigrecurrence}
is valid for any energy spectrum. The main challenges in using Eq.~\eqref{bigrecurrence} are numerical complexity (scaling as $\mathcal{O}(E^2)$) and the necessity to deal with large and small numbers simultaneously\footnote{To evaluate the Eq.~\eqref{bigrecurrence} we have used the C lanuguage  MPFR numerical library for high-precision arithmetic.}. 
Other recurrences of a similar complexity to Eq.~\eqref{bigrecurrence} can be found in~\cite{Weiss1997}.

More generally, the task of finding $\gmne$ and $\gmneex$ is of relevance in many branches of science. It has been shown in~\cite{Bhatia1997Apr}, that the $D-1$ dimensional case is related to the compact lattice animal problem discussed for instance in~\cite{Wu1996}, further related to the Pott model (a generalization of the Ising Hamiltonian). The result can also be used to derive statistical properties of 1D fermions, using the concept of Fermion-Boson transmutation~\cite{Schonhammer1996Sep} (see also~\cite{Kubasiak2005Oct}).
These and further analogies between very different systems have lead to the development of new numerical codes to compute partition functions~\cite{Andrij2010} nothwithstanding that the computation of the number of partitions is a hard numerical challenge belonging to the NP complexity class~\cite{Mertens1998}.

It turns out then, that the microcanonical ensemble, although conceptually the simplest one and directly related to many mathematical problems, is also the most difficult ensemble to use.

\subsubsection{Canonical (CN)}
\label{sec:CN}

Now we turn to the situation when the system of interest can exchange energy, but not particles, with a heat-bath  (a thermostat) of a constant temperature $T$. In equilibrium the system's temperature is obviously the same as the  thermostat but its energy fluctuates.
The ensemble of microstates corresponding to such an arrangement is known as the canonical ensemble~(CN) with the partition function defined as:
\begin{equation}
    {\cal Z}(\beta,N)=\sum_E e^{-\beta E }\gmne,
    \label{eq:cn}
\end{equation}
where states contributing to a given microcanonical energy $E$ are weighted with a Boltzmann factor $e^{-\beta E}$, with $\beta=1/k_B T$ and Boltzmann constant $k_B$.

\begin{figure}
    \centering
    \includegraphics[width=\linewidth]{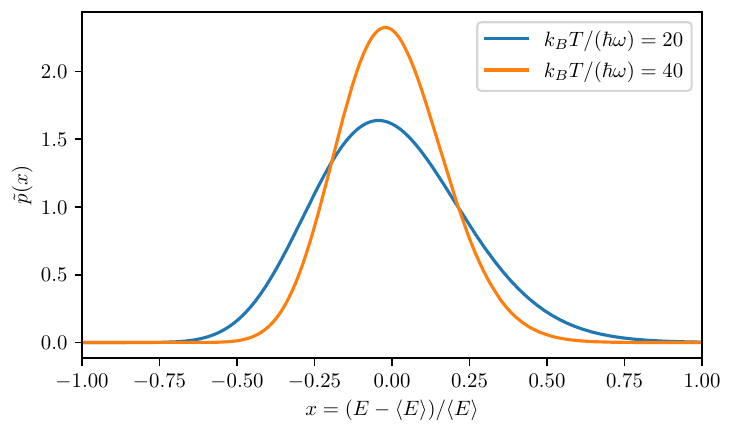}
    \caption{\label{fig:pE} The probability distribution of microstates with energy $E$ and $N=100$ particles, $\mathcal{P}(E)=\Gamma e^{-\beta E}/\mathcal{Z}$, in a CN ensemble in a 1D harmonic trap given by Eq.~\eqref{eq:cn}. The figure illustrates the shrinking relative width of the energy distribution, shown using the normalised probability distribution $\tilde{p}(x)=\langle E\rangle\, \mathcal{P}(\,E(x)\,)$ 
    in the relative variable $x=(E-\langle E\rangle)/\langle E\rangle$.
    }
    \label{fig:pE-micro-cano}
\end{figure}

The Boltzmann factor decreases exponentially with energy, while the microcanonical partition function $\gmne$ grows exponentially. The product of the two has a maximum around the mean energy 
\begin{equation}
\label{eq:mean_E}
    \meanv{E}= -\frac{\partial}{\partial \beta}\ln {\cal Z}(\beta,N),
\end{equation}
as illustrated in Fig.~\ref{fig:pE-micro-cano}.
Therefore, the sum in Eq.~\eqref{eq:cn}, can be approximated by the integral:
\begin{equation}
    {\cal Z}(\beta,N) \approx \ e^{\ln \Gamma(\meanv{E},N) -  \meanv{E}\beta}\int {\rm d}E\, e^{-\frac{(E-\meanv{E})^2}{2\fluct{E}}},
    \label{eq:Xi_int}
\end{equation}
where  $(\fluct{E})^{-1}=(\partial^2/\partial E^2) \ln {\Gamma}({\langle  E\rangle})>0 $ is a squared width of the maximum of the energy distribution.
If, in the limit of $N,E \to \infty$, the width of the distribution scales as $\fluct{E} \sim {\cal O}(\langle E \rangle)$, then:
\begin{equation}
\label{eq:mc_cn_eq}
    \frac{\sqrt{\fluct{E}}}{\langle E \rangle} \sim \frac{1}{\sqrt{\langle E \rangle}}.
\end{equation}
If this condition is met, the ensembles are considered to be \underline{equivalent} (having in mind the thermodynamic limit). 
The criterion Eq.~\eqref{eq:mc_cn_eq}, is met if $\ln \gmne$ is an extensive function~\cite{Huang_1987}. Then the whole sum in Eq.~\eqref{eq:cn} is dominated by a single term,  the one which corresponds to the mean energy given by Eq.~\eqref{eq:mean_E}.
Taking the logarithm of both sides of Eq. (\ref{eq:Xi_int}) 
 and maintaining the dominant terms, one finds that the  partition functions of CN and MC ensembles are   related: 
\begin{eqnarray}
\label{Z_to_G}
   \ln  \Gamma(E,N)  \approx  \ln {\cal Z}(\beta,N) + E \beta, 
\end{eqnarray}
where the canonical mean energy $\langle E\rangle$ equals the microcanonical energy, $\langle E \rangle = E$.
In the above equation, $\beta$ is not an independent variable, but  depends on the energy according to  Eq.(\ref{eq:mean_E}), $\beta=\beta(E,N)$. Eq.~\eqref{Z_to_G} is then the Legendre transformation converting a function  $\ln {\cal Z}(\beta,N) $ of variable $\beta$ into a function $\ln \Gamma(E,N)$ that depends on the  conjugate quantity,  $E=-\partial (\ln{\cal Z})/\partial \beta$.
Different discussions for the equivalence between ensembles can be found in~\cite{Pathria1972}.

A compact expression for the canonical partition function of an ideal gas can be obtained from the following form
\begin{equation}
    \label{eqkp:ZN}
    {\cal Z}(\beta, N)=\sum_{N_1=0}^\infty 
    \sum_{N_2=0}^\infty \cdots \,
    e^{-\beta E} \, \delta_{\sum_j\!N_j, N},
\end{equation}
where the Kronecker delta function 
enforces the condition  of having exactly $N$ atoms in the system, i.e. $N = \sum_{j=0}^{\infty}N_j$ with populations $N_j$ of 
for the $j$th state. Since the gas is non-interacting, therefore the total energy $E$ equals $\sum_j \epsilon_j N_j$, with the single-particle energies $\epsilon_j$. 
Introducing an integral representation
of the delta function, $\delta_{a,b}=\frac{1}{2\pi} \int_0^{2 \pi} e^{i x(a-b)} dx$, one can calculate all sums over $N_j$:
\begin{equation}
    {\cal Z} ( {\beta}, N)= \frac{1}{2 \pi} \int dx e^{-i x N} \prod_{j=1}^\infty \frac{1}{1-e^{-(\beta \epsilon_j + i x)}},
\end{equation}
that simplify further due to the residue theorem for a complex variable $z=e^{i x}$:
\begin{equation}
    {\cal Z} (\beta, N)=\sum_{j=1}^\infty 
    e^{-\beta \epsilon_j (N-1)} \prod_{l\ne j}\frac{1}{1-e^{-\beta (\epsilon_l - \epsilon_j)}}.
\end{equation}
This final formula for the partition function~\eqref{eqkp:ZN} applies to any spectrum provided it is non-degenerate. 
A useful quantity is the partition function for the system of $N$ atoms among which exactly $\nex$ are in the excited states. It can be computed similarly to the derivation presented above but with the Kronecker delta imposing the fixed number $\nex$ of excited bosons:
\begin{equation}
    \label{eqkp:Zex}
    {\cal Z}_{\rm ex}(\beta,\nex)=\sum_{j=1}^\infty 
    e^{-\beta \epsilon_j (N_{\rm ex}-1)} \prod_{l\ne j}\frac{1}{1-e^{-\beta (\epsilon_l - \epsilon_j)}}.
\end{equation}
With this quantity one can compute the probability distribution  $\mathcal{P} (\beta, N_{\rm ex}) =  {\cal Z}_{\rm ex}(\beta, \nex )/{\cal Z}(\beta, N)$~\cite{PhysRevA.79.033631}, and as a consequence also 
expectation values of the number of condensed and excited bosons, as well as their fluctuations. In this special case of a 1D harmonic oscillator, partition functions can be expressed in simpler form:
\begin{eqnarray}
    {\cal Z} (\beta, N) &=& \prod_{j=1}^N\frac{1}{1-e^{-\beta j }} \quad \text{(for 1D h. o.)}\\
    {\cal Z}_{\rm ex} (\beta, \nex) &=& e^{-\beta \nex}\prod_{j=1}^{\nex}\frac{1}{1-e^{-\beta j }} \quad \text{(for 1D h. o.)}
\end{eqnarray}
The same approach can also be generalized to a 1D box 
with periodic boundary conditions under careful consideration of the double degeneracy of each excited state~\cite{Kruk2020}. 

Another procedure for obtaining the canonical partition function uses directly the definition~\eqref{eq:cn} with $\gmne$ obtained via methods described in  Sec.~\ref{sec:MC}. 
Yet another, quite efficient method to obtain ${\cal Z}(\beta, N)$ follows the 
 recurrence relation (see the Appendix of~\cite{Weiss1997} for a quick proof)
\begin{equation}
{\cal Z}(\beta, N) = \sum_{n=1}^N {\cal Z}(n\beta, 1)\,{\cal Z}(\beta, N-n),
\label{eq:weiss-wilkens-recurrence}
\end{equation}
that connects the ``target'' partition function with a smaller number of atoms ${\cal Z}(\beta, N-n)$, and a single particle partition for a low temperature ${\cal Z}(n\beta,\,1)$. Finally ${\cal Z}(\beta, N)$ can be computed from the grand canonical partition function using contour integrals~\cite{Politzer1996} as introduced in the next section.

\subsubsection{Grand Canonical (GC)
\label{sec:GC}}
Thirdly, we consider a system in contact with a reservoir of both energy and particles. We assume that the reservoir has a constant temperature $T$, and particles can flow back and forth between the system and the reservoir. The energy cost of adding a particle to the system is given by the negative chemical potential $\mu$, which is used to define the fugacity $z=e^{\beta \mu}$.  

\begin{figure}
    \centering
    \includegraphics[width=\linewidth]{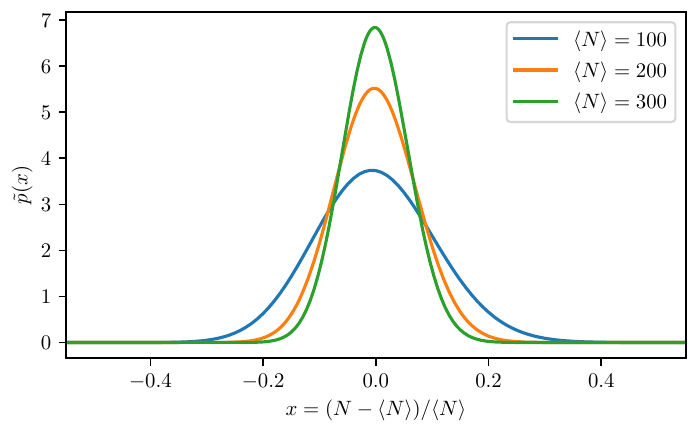}
    \caption{The narrowing of the probability distribution of microstates with $N$ particles, $\mathcal{P}(N)=z^N\mathcal{Z}/\Xi$, in a 3D harmonic trap given by Eq.~\eqref{eq:gk}. We illustrate the shrinking particle number distribution $\tilde{p}(x)$ width (relative to the mean particle number) in the variable $x=(N-\langle N\rangle)/\langle N\rangle$, ${\tilde{p}(x)=\langle N\rangle \mathcal{P}(N(x))}$ for constant, uncondensed, $T/T_c>1$. The chemical potential $\mu$ is fixed by the choice of $\langle N\rangle$.}
    \label{fig:fig3}
\end{figure}

The grand canonical (GC) description does not introduce any constraints on the energy nor on the number of particles. All values are allowed, but Gibbs statistical weights $e^{\eta(\mu N- E)}=z^Ne^{-\beta E}$ are introduced instead. The GC partition function  $\Xi(\beta,z)$, that determines the probability distribution in the space of all microstates with an arbitrary energy and arbitrary number of particles,  is defined as: 
\begin{equation}
    \Xi(\beta,z)=\sum_N z^N{\cal Z}(\beta,N),
    \label{eq:gk}
\end{equation}
where each term in the above sum is proportional to the probability of having $N$ particles in the ensemble. The mean number of particles 
is controlled  by the fugacity $z$, and in the thermodynamic limit:
\begin{equation}
    \langle N\rangle=\frac{\partial }{\partial (\beta \mu)} \ln \Xi (\beta,\mu) =z \frac{\partial }{\partial z} \ln \Xi (\beta,z).
    \label{n_av}
\end{equation}
The entire distribution has a maximum  at $N=\langle N\rangle $, if the value of the second derivative,  $(\partial^2/\partial (\beta \mu)^2) \ln {\cal Z}(\langle N\rangle )$ is negative, what is the usual case, see~\cite{Huang_1987}. Then the derivative can be interpreted as $(\fluct{N})^{-1}$. Moreover, because $(\fluct{N})$ is extensive,
the criterion of equivalence between the GC and CN ensembles takes the form:
\begin{equation}
    \frac{\sqrt{\fluct{N}}}{\langle N \rangle}=\frac{1}{\sqrt{\langle N \rangle}},
\end{equation}
i.e. in the limit of $\langle N\rangle  \to \infty$ the probability distribution becomes a delta-like function focused on the mean number of particles, a trend shown in Fig.~\ref{fig:fig3}. 

In such a case the canonical and grand canonical partition functions are related. The transformation  from ${\cal Z}(\beta,N)$ into $\Xi(\beta,z)$ is the Legendre  transformation: 
\begin{equation}
\label{legrenge_Z}
    \ln {\cal Z}(\beta,N) \approx \ln \Xi(\beta,z) - N \ln z.
\end{equation}
As we will demonstrate in the next section, a Bose-Einstein condensate is a system in which this equivalence between different statistical ensembles is violated.

In contrast to the MC and CN formalisms, where determining partition functions analytically is possible only for a few special cases, obtaining the GC partition function is much simpler. This is because detailed constraints on the number of particles and energy are relaxed. 

Summation over all single-particle energies $\epsilon_j$ in the spectrum can be substituted by summation over all one-particle states and corresponding occupation numbers $N_j$. 
For bosonic particles the 
$N_j$ can take any integer value from zero to infinity. Had we considered fermionic particles, $N_j$ could take only two values, zero or one. 
The energy of the system is 
$E=\sum_j N_j \epsilon_j$ and $N$ 
is the sum of occupations of all states. 

According to Eq.~\eqref{eq:gk} we have:
\begin{eqnarray}
\label{Xi_2}
    &&\Xi(T,\mu) =  \sum_{N=0}^\infty \sum_{\{N_j\}_N}  e^{-\sum_j  N_j\beta (\epsilon_j-\mu)}  = \\
     && =  \left(\sum_{N_0=0}^\infty  \ldots \right) \ldots \left(\sum_{N_j=0}^\infty 
     e^{- N_j \beta (\epsilon_j-\mu)} \right)  \ldots, \nonumber
\end{eqnarray}
where by $\{N_j\}_N$ we denote all sets of occupation numbers satisfying condition $\sum_j N_j = N$.
From the above equation, by noticing that the terms in brackets are geometric series, we get:
\begin{equation}
    \ln {\Xi}(\beta, z)=-\sum_{\epsilon_j} \ddj \ln\left( 1-ze^{- \beta \epsilon_j} \right),
\end{equation}
where now summation goes over all energies and $\ddj$ is a factor accounting for their degeneracy~\cite{Huang_1987}.

We now focus on the most important case experimentally, the
ideal gas 
in a 3D isotropic harmonic oscillator potential of frequency $\omega$. 
As in Sec. \ref{sec:MC} 
we shift the spectrum by the lowest energy (here $3\hbar\omega/2$) such that the ground state energy is $0$, and we use oscillatory units for energy $\ve_0 = \hbar \omega$,
dimensionless temperature $k_BT/\ve_0$.

The one-particle eigenenergies remain integers 
$\epsilon_j=j$ and  
the degeneracy of the energy state $\epsilon_j$ is $\ddj=(j+2)(j+1)/2$ (or $\ddj=j+1$ in 2D 
and $\ddj=1$ in the 1D case). 

The asymptotic expression for the GC partition function $\Xi$ can be easily found. Using the Euler-MacLaurin formula to substitute summation over  a discrete energy spectrum by integration and by noticing that $\ln(1-x)=-\sum_{l=0}^\infty \frac{x^l}{l}$ we have:
\begin{equation}
    \ln \Xi(\beta,z) =  -\ln(1-z)+ \ln \Xi_{\rm ex}(\beta,z),
\end{equation}
where the term $\ln \Xi_{\rm ex}(\beta,z)$ is the contribution from excited states:
\begin{eqnarray}
\label{Xi_ex}
\ln \Xi_{\rm ex}(\beta,z) &\approx& \sum_{l=0}^\infty \frac{z^l}{2l} 
\int_0^\infty\!\!d\vej \, e^{-l \beta \vej} (\vej + 2)(\vej + 1)  \nonumber \\
&\approx&\frac{1}{\beta^3}\, g_{4}(z) ,
\end{eqnarray}
The Bose  function $g_\alpha(z)$ 
is defined as 
\begin{equation}\label{boseg}
g_\alpha(z)=\sum_{j=1}^\infty \frac{z^j}{j^\alpha},
\end{equation}
and the series is convergent for $0\le z \le 1 $ (if $\alpha >1$). That is, over the physical range of variation of the fugacity 
$z=e^{\beta\mu}$  because the chemical potential for an ideal Bose gas is not larger than the one-particle ground state energy. If $\alpha > 1$, the values of $g_\alpha(z)$ at $z=1$ are  equal to the Riemann zeta function $g_\alpha(1)=\zeta(\alpha)$ and are finite. If $\alpha=1$ then $g_1(z)=-\ln(1-z)$, and  has a logarithmic divergence  at $z=1$. For  $\alpha<1$ the divergence is stronger. It is straightforward to check also that 
\begin{equation}\label{zdzg}
z\frac{d}{dz}g_\alpha(z)=g_{\alpha-1}(z).
\end{equation}

In (\ref{Xi_ex}) we omitted terms proportional to $1/\beta^2$ and $1/\beta$ because, as it will become clear soon, we are interested in the small $\beta$ (i.e. large temperature) behaviour of $\ln \Xi(\beta,z)$. 
Combining the two contributions, from the ground and excited states we have:
\begin{equation}
\label{GCPF}
    \ln \Xi(\beta,z)  \approx -\ln(1-z)+\frac{1}{\beta^3}\, g_{4}(z).
\end{equation}
The temperature and mean energy of the system are related:
\begin{equation}
\label{E-T}
    \wb{E}=-\frac{\partial}{\partial \beta} \ln \Xi(\beta,z) = 3  \frac{g_4(z)}{\beta^4} \sim T^4,
\end{equation}
which reminds one of the Stefan-Boltzmann law.

\subsection{Bose-Einstein condensation}
\label{section_BEC}

The textbook statistical description of a Bose-Einstein condensation is traditionally made within the GC ensemble following the very first ideas of A.~Einstein~\cite{Einstein1924, Huang_1987},
and considers the thermodynamic limit of a $D$ dimensional system in which the system size does not affect the intensive quantities, 
\begin{equation}
\langle N\rangle\to \infty, \,\, \beta \to 0, \,{\rm but}\,\,  \langle N\rangle\beta^D = {\rm const.}     
\label{eq:thermlimit}
\end{equation}
In this formalism, and focusing on the
ideal gas in a 3D isotropic harmonic oscillator potential as considered in \ref{sec:GC}, 
the mean number of particles $\langle N\rangle$ in the system depends on the temperature and the chemical potential/fugacity,
\begin{equation}
    \langle N \rangle= \frac{z}{1-z}+\frac{1}{\beta^3} \, g_3(z),
\label{N_har}
\end{equation}
see Eq.(\ref{n_av}).
The two terms on the right-hand side have physical interpretations: the mean occupation of the ground state and the mean occupation of all excited states, respectively:
\begin{eqnarray}
    \langle N_0\rangle &=& \frac{z}{1-z}, \label{N0z}\\
    \langle N_{\rm ex}\rangle &=&\frac{g_3(z)}{\beta^3},\label{Nexz}
\end{eqnarray}
For a given inverse temperature $\beta$, the mean number of particles in excited states is limited by the value 
\eq{Nc}{
\langle N_{\rm ex}\rangle \le \frac{g_3(1)}{\beta^3} = \frac{\zeta(3)}{\beta^3}=N_c.
}
Alternatively one can define a `critical' value of $\beta=\beta_c$ as the temperature when maximal `capacity' of excited states $N_c$ is equal to the mean number of all particles, $\langle N\rangle$, in the system: 
\begin{equation}
\label{beta_C}
    \beta_c=\frac{1}{T_c}= \left(\frac{\zeta(3)}{\langle N\rangle }\right)^{1/3}.
\end{equation}
Combining Eq.~\eqref{E-T} and Eq.~\eqref{beta_C} the mean energy of the system at the Bose-Einstein transition can be found:
\begin{eqnarray}
\label{EC}
E_c=3\frac{\zeta(4)}{\beta_c^4} = 3 \zeta(4) \left(\frac{\langle N\rangle }{\zeta(3)}\right)^{4/3}.    
\end{eqnarray}

It is convenient to rewrite equation Eq.~\eqref{N_har} as: 
\begin{equation}
    \langle N\rangle \beta^3= \langle N_0\rangle (z)\beta^3 + g_3(z).
\label{N_equation}
\end{equation}
Equation~\eqref{N_equation} can be thought of as the equation determining a value of $z$ for a given mean number of particles $\langle N\rangle $ (which now becomes a paremeter of the system)and a temperature. 
In the thermodynamic limit (\ref{eq:thermlimit}) 
and above critical temperature the first term on the right-hand side of Eq.(\ref{N_equation}) vanishes,
$\lim_{\beta \rightarrow 0}\langle N_0\rangle (z)\beta^3=0$, because the fugacity $z$ found from: 
\begin{equation}
\label{beta3}
g_3(z)=\langle N\rangle \beta^3,    
\end{equation}
is  $z<1$, and consequently $\langle N_0\rangle $ from Eq.~\eqref{N0z} is finite (i.e. negligible in the thermodynamic limit).
Below the critical temperature, i.e. for $\beta > \beta_c$, the fugacity must remain fixed at $z=1$, in order to keep $\langle N\rangle \beta^3$ finite ($g_3(z>1)$ diverges) and one obtains a non-zero mean fraction,
$\lim_{\langle N\rangle \rightarrow \infty}\langle N_0\rangle /\langle N\rangle  > 0$, of particles in the ground state. 
This  effect of sudden accumulation  of a finite fraction of atoms in the ground state, usually observed while cooling the system down below critical temperature,  is known as Bose-Einstein condensation.  
Equation (\ref{N_equation}) allows one to find 
the fraction of condensed particles: 
\begin{equation}
        \frac{\langle N_0\rangle}{\langle N\rangle} =1-\left(\frac{T}{T_c}\right)^{3}.
\label{cond_frac_har}
\end{equation}
The phase transition to the Bose-Einstein condensate is a direct consequence of a `finite capacity' of excited states at fixed temperature. The above ``textbook'' reasoning however, is loosely based on the assumption that the mean number of particles in the system remains constant or well controlled in this formulation also when 
temperature is decreased below the critical value. A potential problem with this,  which we will return to in Sec.~\ref{CAT}, is that $\langle N\rangle $ is extensive and not an original control parameter of the model
formulated in \eqref{eq:gk} and \eqref{Xi_2}, which was given in terms of only intensive parameters $z$ and $T$.
This leads to the following paradox: below $T_c$ we found that fugacity is constant $z=1$, therefore it cannot serve as a control knob in the model for the average number of particles, and the two-dimensional parameter space of systems with differing $\langle N\rangle $ and $T$ is left underconstrained with only one control parameter $T$.

Bose-Einstein condensation depends on spatial dimensionality and on the shape of external trapping potential. This is because the phenomenon is sensitive to the density of states as we saw in Eq.~\eqref{Xi_ex}. 
Calculations analogous to those presented above, allow one to find analytic expressions for the grand canonical partition function in the thermodynamic limit for dimensionalities $D=2$ and $D=1$: 
\begin{equation}\label{12dlogXi}
    \ln \Xi^{(D)}(\beta,z)  \approx -\ln(1-z)+\frac{1}{\beta^D}\, g_{D+1}(z).
\end{equation}
Based on this result, the relation between the fugacity and mean number of particles in the system, takes the form (compare Eq.~\eqref{N_equation}):
\begin{equation}
\label{N_D}
\langle N\rangle \beta^D  = g_D(z) + \langle N_0\rangle  \beta^D.
\end{equation}
If $\langle N_0\rangle $ is finite, the second term in the right-hand side of the above equation can be neglected when passing to the thermodynamic limit~\eqref{eq:thermlimit}. 
Evidently occupation of the ground state $\langle N_0\rangle $ is finite in the limit of $N \to \infty$, hence $\langle N_0\rangle /\langle N\rangle $ is negligible, as long as there exists a $z$ solving the equation 
\begin{equation}\label{N_D_above}
    \langle N\rangle \beta^D=g_D(z)
\end{equation}
analogous to Eq.(\ref{beta3}).
In two dimensions, $g_2(z)$ is limited by the value $g_2(1)=\zeta(2)=\pi^2/6$. If $\langle N\rangle \beta^2> \zeta(2)$, there is no solution of \eqref{N_D_above}. For a fixed value of $\langle N\rangle $, there exists a critical temperature above which Eq.~\eqref{N_D_above} is satisfied and all particles accumulate in the excited states:
\begin{equation}\label{b2}
    \beta_c^{D=2}=\frac{1}{T_c^{D=2}}= \left(\frac{\zeta(2)}{\langle N\rangle }\right)^{1/2}.
\end{equation}
If temperature decreases below the critical one and assuming that the mean number of particles in the system does not change, then because the capacity of excited states saturates at $N^{D=2}_c(T)=\zeta(2)/\beta^2$, the equation Eq.(\ref{N_D}) can be solved with $z=1$, and \eqref{b2} giving a mean fraction of ground state particles that is finite:
\begin{equation}
   \frac{\langle N_0\rangle}{\langle N\rangle} =1-\left(\frac{T}{T_c^{D=2}}\right)^2.
\end{equation}
This demonstrates the Bose-Einstein condensation of the ideal gas trapped in a harmonic potential in two dimensions.  Interestingly, the same 
transition is not present in the uniform system trapped in a  
box (or in periodic boundary conditions).

In the presence of interactions, there the Kosterlitz-Thouless transition takes place instead, as measured for cold atoms in~\cite{Hadzibabic2006Jun}.

In the one-dimensional trap, the equation determining the fugacity $z$ for a given $\langle N\rangle $ is:
\begin{equation}
\label{beta1}
    \langle N\rangle \beta = \langle N_0\rangle  \beta -  \ln(1-z).
\end{equation}
In the thermodynamic limit (\ref{eq:thermlimit})
the first term to the right-hand side vanishes if $\langle N_0\rangle $ is finite. 

As the logarithm is not bound, the fugacity can be found for any value of $\langle N\rangle \beta$, $z=1-e^{-\langle N\rangle \beta}$, without any need to place a finite fraction of particles, $\langle N_0\rangle /\langle N\rangle $, in the ground state at nonzero temperature.  Therefore, there is no Bose-Einstein phase transition in an ideal gas trapped in a 1D harmonic potential. 

However, because the logarithm grows very slowly,  if a system is finite there is still quite an appreciable occupation of the ground state, even at significantly nonzero temperature~\cite{Ketterle96}. This occupation gradually increases as the temperature approaches zero. The characteristic value of the fugacity, at which  $\langle N_0\rangle $ is of the order of $\langle N\rangle $, $z/(1-z)=\langle N\rangle $, equals $z=1-1/\langle N\rangle $. Then  Eq.~(\ref{beta1})  allows us to define (neglecting the first term in the right-hand side) a characteristic temperature at which the population of the ground state becomes significant \footnote{A similar characteristic transition temperature was defined in~\cite{Ketterle96} via $\langle N\rangle \beta=\ln(2/\beta)$.}:
\begin{equation}
\label{T1}
    \beta_c^{D=1} = \frac{1}{T_c^{D=1}} = \frac{\ln \langle N\rangle }{\langle N\rangle }.
\end{equation}
By combining Eq.(\ref{beta1}) and Eq.(\ref{T1}) the temperature dependence of the ground state occupation can be estimated: 
\begin{eqnarray}
  \frac{\langle N_0\rangle}{\langle N\rangle}  = 1 -\frac{T}{T_c^{D=1}}.
\end{eqnarray}
The above picture is strongly modified by interactions. 
In~\cite{PhysRevLett.85.3745}, 
different regimes of quantum degeneracy of the 1D harmonically trapped system are identified: a true BEC condensate, a quasicondensate, and a gas of impenetrable bosons — the Tonks-Girardeau gas.

\subsection{Fluctuations catastrophe}
\label{CAT}

The GC description of Bose-Einstein condensation has a serious drawback, which leads to the conclusion that the GC  formalism fails below the Bose-Einstein condensation temperature for known systems in the thermodynamic limit~\cite{Ziff1977}.

The fugacity in this limit is equal to one, $z=1$, thus no longer plays the role of a parameter controlling the mean number of particles. 
All canonical ensembles with different $N$ contribute then to the GC partition function with the same weights, because we have $z^N=1^N=1$. 
Therefore, the assumption that the number of particles (or mean number of particles) at and below the critical temperature remains well controlled 
while decreasing the temperature at constant $z=1$,  which is used on \eqref{N_equation} to get \eqref{cond_frac_har}, is not well justified. 

This hidden but dubious assumption that one can swap around $z$ and $\langle N\rangle $ in the GCE is fundamental for the textbook formulation of Bose-Einstein condensation. It lies behind Eq.~(\ref{cond_frac_har}), despite the fact that, when in the thermodynamic limit, the GC formalism does not provide any knobs to control the value of $\langle N\rangle $ after reaching or crossing the critical temperature. 

Fluctuations of the ground state occupation $\fluct{N_0} \equiv \langle N_0^2 \rangle - \langle N_0 \rangle^2$, serve as the magnifying glass for the problems of the grand canonical description of the Bose-Einstein condensation, and expose the poor control over $N$ or $\langle N\rangle$ in the model. 
From  the definition of the grand canonical partition function Eq.~\eqref{Xi_2} it follows that the mean occupation $\meanv{N_i}$ of one-particle  states of energy $\epsilon_i$,  and its fluctuations (squared) $\fluct{N_i}$  are 
(compare Eq.\eqref{n_av}):
\begin{eqnarray}
\label{mean n_i}
 \meanv{N_i} & = & -\frac{\partial \ln \Xi(\beta,\mu)}{\beta\partial \epsilon_i}, \\
 \label{fluc n_i}
 \fluct{N_i} & = & \frac{\partial^2 \ln \Xi(\beta,\mu)}{\beta^2\partial \epsilon_i^2}=  \meanv{N_i}\left(1+\meanv{N_i}\right).\quad
\end{eqnarray}
Eq.~\eqref{fluc n_i} is valid for all temperatures.  If applied to the ground state occupation below the condensation temperature where $\langle N_0\rangle \sim \langle N\rangle \gg 1$ we find that:
\begin{equation}
\label{flu_n_0}
    \fluct{N_0} = \langle N_0\rangle(\langle N_0\rangle+1) \approx \langle N_0\rangle^2 \sim \langle N\rangle^2.
\end{equation}
The fluctuations $\delta N_0$  
are comparable 
to the total (mean) number of particles in the system.  The result is extravagantly large and 
in fact conceptually suspicious, as has been extensively discussed in the literature~\cite{Ziff1977,Grossmann97arxiv,Johnston70,Yukalov04}.

Fluctuations of the total number of particles can be obtained from: 
\begin{equation}
 \fluct{N} =  \frac{\partial^2 \ln \Xi(\beta,\mu)}{\beta^2\partial \mu^2} = \left(z\frac{\partial}{\partial z}\right)^2   \ln \Xi(\beta,z),
\end{equation}
and consequently with division of the GC partition function into the ground state contribution and the excited particles term, Eq.(\ref{GCPF}), the two subsystems contribute:
\begin{equation}
\label{Fluc_tot}
 \fluct{N} = \fluct{N_0} + \fluct{ \nex },
\end{equation}
where $\fluct{N_0}$ is given by Eq.(\ref{flu_n_0}) 
while  the second term  of  Eq.(\ref{Fluc_tot}) gives fluctuations of the excited subsystem, the thermal cloud: 
\begin{equation}
  \fluct{ \nex }=  \left(z\frac{\partial}{\partial z}\right)^2   \ln \Xi_{\rm ex}(\beta,z) = \frac{g_2(z)}{\beta^3}.
\end{equation}
Below the condensation temperature $T<T_c$ the fugacity equals to one, $z=1$, and by substitution of quantities from Sec.~\ref{section_BEC} fluctuations of the number of particles in the thermal cloud  can be written as:
\begin{equation}
\label{fluc_thermal}
  \fluct{\nex} = (\langle N\rangle-\langle N_0\rangle)\frac{\zeta(2)}{\zeta(3)}=\langle\nex\rangle \frac{\zeta(2)}{\zeta(3)}.  
\end{equation}
It becomes clear now, that the total number of particles in the system fluctuates excessively only because of the enormous fluctuations of the ground state, illustrated in Fig.~\ref{fig:graphical-abstract} in green. 
The problem fluctuations are solely related to the fact, that while the total number of particles in the system described by the GC ensemble (GCE) is free to take on any value there is no associated energy cost for entering or leaving the ground state. 
Having seen this, we can revisit the conceptual problems of the GCE below $T_c$ from other angles than the poor conditioning of $N,\langle N\rangle $ on $z=1$. 
Another conceptually dubious element there is that such strongly fluctuating $\delta N\sim \langle N\rangle$ 
makes it difficult to define the ``system'' well in the presence of a particle reservoir. Separation of system from reservoir becomes then mostly an exercise in arbitrary labeling.
For example,  Ref.~\cite{Ziff1977} points out and earlier~\cite{Johnston70} also intimated, that in the limit $V\to\infty$ of a subsystem included inside an even larger $V'\gg V$ volume of an ideal gas whose remaining parts are the ``reservoir'', the properties of the smaller volume $V$ subsystem tend to whenever condensation is present those of a canonical ensemble not a GCE. This is ascribed to the off-diagonal long-range order. Hence the often espoused view that the GCE at least describes properly a ``small section'' of a larger ideal gas is a fallacy below $T_c$.

From another angle, it has been noted that in CN and MC ensembles relevant physical quantities such as condensate fluctuations $\delta N_0(T)$ become independent of $N$ as $N$ grows, apart from $N$ determining the global condensation temperature. This makes the introduction of a particle reservoir at a given $T\ll T_c$ basically irrelevant, and any resulting fluctuations of quantities that are calculated can again be suspected to be merely a consequence of this spurious labeling problem between system and reservoir particles. 

Yet another view is given in a recent work~\cite{Yukalov24}, which proposes removal of the GCE fluctuation catastrophe via gauge symmetry breaking arguments. There it is argued that no condensate fluctuations take place in a GCE because of an imperative for symmetry breaking, and instead any remaining large particle fluctuations lie in excited modes. The need for symmetry breaking was supported e.g. by the result that a small symmetry breaking parameter $\varepsilon$ added to an ideal gas Hamiltonian provides the leading contribution to the mean ground state occupation $\langle N_0\rangle \approx \varepsilon^2V/\mu^2-T/\mu + \dots$ in finite systems. i.e. the  better known chemical potential contribution $z/(1-z)\approx -T/\mu$ is sub-leading \cite{Yukalov24}. 

The discussion presented so far reveals that a straightforward 
description of the ideal Bose gas according to only classical statistical mechanics fails at sufficiently low temperatures. The assumption about the equivalence of ensembles, a pillar of equilibrium statistical mechanics, is no longer true. The main reason is that below the critical temperature, most particles occupy the single quantum state of zero energy, thus energy is no longer an extensive quantity. The chemical potential equals the ground state energy and does not control the number of particles in the system which leads to divergent 
fluctuations. Therefore, there is a need to determine a formalism of Bose-Einstein condensation for the ideal Bose gas with a well-defined total number of particles.

The standard solution the problems of the GCE below $T_c$ has been to employ a formalism in which the total number of particles is fixed -- the canonical (CN) and microcanonical (MC) descriptions. However, the constraint on the number of particles adds significant complexity to the problem. This challenge was addressed by H. David Politzer in~\cite{Politzer1996}, where the CN arrangement is assumed. In this scenario,  the temperature is set by the heat reservoir, but the number of particles is constant, leading to explicitly imposed vanishing of its fluctuations, $\fluct{N} =0$. 

Consequently,  the fluctuations of the condensate, $\delta^2_{CN} N_0$,  can be deduced from the fluctuations of the thermal cloud:
\begin{equation}
    \delta^2_{CN} N_0=\delta^2_{CN} N_{\rm ex},
\end{equation}
where, to stress that we address the CN formalism, labels are added. Assuming that $\Xi_{\rm ex}$ still provides a correct description of the thermal cloud which is in equilibrium with the ground state that acts as a reservoir of particles,  the fluctuations of the number of particles in the excited subsystem can be obtained from the scheme of the grand canonical ensemble by Eq.(\ref{fluc_thermal}). These when combined with Eq.(\ref{cond_frac_har}) give:
\begin{equation}
\label{CN_fluc}
    \delta^2_{CN} N_0 = N \, \frac{\zeta(2)}{\zeta(3)} \left( \frac{T}{T_c} \right)^3.
\end{equation}
Note that the mean total number of particles appears only to compensate the $\langle N\rangle ^{1/3}$ appearing in the condensation temperature $T_c$ from \eqref{beta_C}. 
This remarkably straightforward result demonstrates that the relative fluctuations of the condensate are normal, i.e. $\sqrt{\delta^2_{CN} N_0}/N \sim 1/\sqrt{N}$. Its scaling was previously shown in~\cite{Gajda97}, and (\ref{CN_fluc}) can be compared to the orange numerical curve in Fig.~\ref{fig:graphical-abstract}. 
 
Finally, it is worth remembering that 
a GC formulation may remain useful for finite-size systems under other physical conditions. For example, the GC is valid in systems such as the so-called photon condensates~\cite{Weiss16} in which there are real fluctuations of the number of particles in a system whose extent is well defined by other physical effects. Another case is the ``grand'' canonical formulation of~\cite{Kocharovsky2006} in which $\langle N_0\rangle$ 
is made to determine an effective fugacity $z_{\rm eff}=\langle N_0\rangle /(\langle N_0\rangle +1)$ consistent with \eqref{N0z}, which $z$ then enters GC expressions for calculation of observables.
Finally, it is believed that for the interacting gas, the equivalence of different ensembles can be restored.

\subsection{The Maxwell Demon (MD) rescue ensemble}
\label{sec:maxwelldeamon}
In this section we explore in depth the original idea of~\cite{Politzer1996},  that the ground state contribution should be excluded from the statistical description, which treats only the excited subsystem of the thermal cloud. Constraints on the total number of particles, necessary for realistic condensation should be imposed externally. 

We focus on the region of energies below the transition to the Bose-Einstein condensation phase, where $E < E_c$ from \eqref{EC} and $z=1$. Guided by the idea of a statistical description of the Bose-Einstein condensate in terms of quantities related to the excited subsystem only, one classifies all microstates based on the abundance of particles in the excited states. The recall that the MC partition function can be organized as:
\begin{equation}
\label{gam}
    \Gamma(E,N)=\sum_{N_{\rm ex}=0}^N \Gamma_{\rm ex}(E,N_{\rm ex}),
\end{equation}
where by $\Gamma_{\rm ex}(E,N_{\rm ex})$ we denote the number of microstates with exactly $N_{\rm ex}$ excited particles, as discussed previously.

It is convenient to define the following partition function:
\begin{equation}
    \label{maxwell}
    \Upsilon(E,z) = \sum_{\nex=0}^\infty z^{\nex} \gmneex.
\end{equation}
Which for  $z=1$, becomes the MC partition in the thermodynamic limit of an infinite number of particles, $\Upsilon(E,z=1)=\Gamma(E,N \to \infty)$.
The  $\Upsilon(E,z)$ provides a description of the excited subsystem exchanging particles with the reservoir of particles in the ground state of zero energy.  This fourth statistical ensemble, introduced in~\cite{Navez1997}, is known as the Maxwell Daemon (MD) ensemble.  Since the transfer of atoms between ground and excited states is not accompanied by any energy transfer, $z=1$, when a particle leaves or enters the thermal cloud, the others have to adjust their energies accordingly.
These unusual correlations invoke the concept of the Maxwell Daemon, a hypothetical entity with knowledge of the energies of all particles, allowing it to control the exchange of atoms between the thermal cloud and the zero-energy condensate, thereby maintaining the total energy constant.

In the thermodynamic limit, the mean number of excited particles, as well as its fluctuations, depend solely on the energy $E$ of the system. Taking the limit $z\to1$ we obtain MC averages $\left< \cdot\right>_{MC}$ for the excited states: 
\begin{eqnarray}
\label{MC_mean1}
    \left< N_{\rm ex}\right>_{MC}&=&z\frac{\partial}{\partial z}  \ln \Upsilon(E,z) \vert_{z=1}, \\
\label{MC_mean2}    
    \delta_{MC}^2N_{\rm ex}&=&\left(z\frac{\partial}{\partial z}\right)^2 \ln \Upsilon(E,z) \vert_{z=1}.
\end{eqnarray}
The fugacity $z$ has to be set to one after differentiation.

The above equations do not depend on the total number of particles. This is because, in the thermodynamic limit, the excited subsystem does not have any means to learn about the occupation of the ground state. The ground state reservoir 
contributes neither to the energy nor to the entropy. Eqs.(\ref{MC_mean1}, \ref{MC_mean2}) are defined in the whole possible energy range from zero to infinity.

To relate these results to a large but finite system with a condensate, we should impose a constraint on $N$. 
This can be done using 
the value of the energy at the condensation transition, $E_c(N)$ from Eq.\eqref{EC}, at which the mean number of excited bosons is equal to the externally known total number of particles:
\begin{equation}
   \left< N_{\rm ex}(E_c)\right>_{MC} \equiv N.
\end{equation}
{The energy $E$ cannot exceed this maximal excitation energy $E_c(N)$, limiting the domain of functions, $\left< N_{\rm ex}(E)\right>_{MC}$, and $\delta_{MC}^2N_{\rm ex}(E)$ to the interval $E \in [0,E_c]$. This also limits the maximal number of excited particles to $N$.
In particular then,
expressions Eq.(\ref{MC_mean1}) and Eq.(\ref{MC_mean2}) are valid for finite systems at energies below and at the critical one. 

The fact that the upper limit of summation in $\Gamma(E,N)$, Eq.~\eqref{gam},
is finite as opposed to $\Upsilon(E,z)$, Eq.~\eqref{maxwell}, where summation runs to infinity, does not play any role 
below the ``restriction temperature'' at which $E/\hbar\omega\to N$, that lies very close to and just below $T_c$ ~\cite{Grossmann1997}. 
The extra terms in Eq.~\eqref{maxwell} are negligible compared to the main contribution to the sum. This is because the relative width of the maximum at the $\Gamma(E,\left< N_{\rm ex}(E_c)\right>_{MC})$ tends to zero in the thermodynamic limit, as illustrated in Fig.~\ref{fig:pE-micro-cano}. 

By imposing a constraint on $N$, the  mean occupation of the condensate as well as its fluctuations can be  related to those of the thermal cloud: 
\begin{eqnarray}
\left< N_0\right>_{MC} &=& N-\left< \nex\right>_{MC}, \\
\delta_{MC}^2N_0 &=& \delta_{MC}^2\nex,
\end{eqnarray}
as announced in Eqns.~\ref{eq:fluct-n0-general}. 
This procedure excludes the condensate mode from the expression \eqref{fluc n_i} for fluctuation of mode occupations which led to physically incorrect results \eqref{flu_n_0}.
Application of the above formalism requires knowledge of the Maxwell Daemon partition function,  given by Eq. (\ref{maxwell}). Directly obtaining it from the definition proves challenging due to the energy constraint. However, relaxing this constraint simplifies the problem.
The partition function of the GC ensemble of excited particles is a generating function of the Maxwell Daemon partition, $\Upsilon(E,z)$:
\begin{eqnarray}
\label{MD_CN}
    \Xi_{\rm ex}(\beta,z) &=& \sum_E e^{-\beta E}\Upsilon(E,z) \nonumber \\ &=& \sum_E \sum_{N_{\rm ex}} e^{-\beta E} z^{N_{\rm ex}}\Gamma(E,N_{\rm ex}).\quad
\end{eqnarray}
The function $\Xi_{\rm ex}(\beta,z) $ can be interpreted as a canonical distribution for the MD ensemble. 
For example, in the case of particles in the 3D harmonic potential this function is given by Eq.~(\ref{Xi_ex}), $\ln \, \Xi_{\rm ex}(\beta,z)=g_4(z)/\beta^3$. 
The mean value of the critical temperature, critical energy or mean occupation of the condensate, are identical in the  GC  formalism discussed in Sec.~\ref{section_BEC}. The CN formalism is based on $\Xi_{\rm ex}(\beta,z)$, however, the fluctuation of the condensate, pathological in the GC ensemble, are normal in the CN formulation with $\Xi_{\rm ex}(\beta,z)$,  Eq.~\eqref{CN_fluc}.  

The MD partition function $\Upsilon$ can be related to $\Xi_{\rm ex}$,  via the Legendre transformation, in the same way as the MC partition $\Gamma$ is related to the CN partition ${\cal Z}$. The terms of the sum in Eq.~(\ref{MD_CN}) have a pronounced  maximum at an energy which can be obtained from:
\begin{equation}
\label{s_point}
    -\frac{\partial}{\partial \beta} \ln \Xi_{\rm ex}(\beta,z) = E,
\end{equation}
and therefore, when the relative width of this maximum goes to zero in the thermodynamic limit,  the sum  in Eq.~(\ref{MD_CN}) is dominated by the value of the largest term, and we have:
\begin{equation}
\label{saddle}
    \ln \Upsilon(E,z)= \ln \Xi_{\rm ex}(\beta,z) +\beta E.
\end{equation}
The inverse temperature $\beta$ can be obtained from Eq.(\ref{s_point}), and thus it can be made dependent on the energy and fugacity $\beta=\beta(E,z)$. 

The early attempts to determine fluctuations of the Bose-Einstein condensate in a 3D harmonic potential,~\cite{Gajda97}, applied two consecutive Legendre transformations. The first one, to obtain the canonical partition function, $\ln {\cal Z}(\beta, N) \approx \ln \Xi(\beta,z)-N\ln z$, by relating fugacity to other parameters, $z=z(N,\beta)$. In the second step, the microcanonical distribution was associated with the canonical one, $\ln \Gamma(E,N) = \ln {\cal Z}(\beta,N) +\beta E$, by replacing $\beta$ by the appropriate value of energy $\beta=\beta(E,N)$. This approach, utilizing the saddle point method twice, turned out to work only for small systems close to the critical temperature, because the singular contribution from the ground state made the grand canonical partition function not well behaved in the thermodynamic limit~\cite{PhysRevE.60.6534}. 

In contrast, Eq.~(\ref{saddle}), and Eq.~(\ref{s_point}) allow one to connect microcanonical fluctuations of the number of excited particles, $\delta_{MC}^2 N_{\rm ex} = (z\tfrac{\partial}{\partial z})^2\ln \Upsilon(E,z)|_{z=1}$, to their canonical counterpart, $\delta_{CN}^2N_{\rm ex} = (z\tfrac{\partial}{\partial z})^2\ln \Xi_{\rm ex}(\beta,z)|_{z=1}$: 
\begin{equation}\label{dNMCCN}
    \delta_{MC}^2N_{\rm ex}=\delta_{CN}^2N_{\rm ex}-\frac{\left< EN_{\rm ex}\right>^2_{CN}}{\delta^2_{CN} E},
\end{equation}
Therefore, the fluctuations in the MC system are always smaller than those in the CN system
because of  nonvanishing correlations between the energy and number of excited particles in the system,  $\left< EN_{\rm ex}\right>^2_{CN}$.

For example, the above general results can be applied to the specific case of particles in 3D harmonic potential. There the equation relating the MC energy to the CN temperature $\beta$ has the form~\eqref{E-T}:
\begin{equation}
    E=3 \frac{g_4(z)}{\beta^4}\quad\to\quad \beta(z,E) = \left(\frac{3g_4(z)}{E}\right)^{1/4}
    \label{betaz}
\end{equation}
Thus the explicit form of $\ln \Upsilon(E,z)$, in the thermodynamic limit, using \eqref{Xi_ex} in \eqref{saddle}, is 
\begin{equation}
    \ln \Upsilon(E,z)= 4 \frac{g_4(z)}{\beta^3}.
\end{equation}
With that in mind, obtaining the mean occupation and fluctuations of the Bose-Einstein condensate in the perfectly isolated system, i.e., in the microcanonical ensemble, is simply a matter of straightforward differentiation and application of the formula for the critical temperature \eqref{beta_C}, just having to remember that $\beta(z)$ as above in \eqref{betaz}. This procedure gives:
\begin{eqnarray}
\label{MC_expressions}
    \left< N_0\right>_{MC}&=& N\left(1- \left(\frac{T}{T_c}\right)^3\right) \\
    \delta_{MC}^2N_0&=& N \left(\frac{\zeta(2)}{\zeta(3)} - \frac{3\zeta(3)}{4\zeta(4)}\right) \left( \frac{T}{T_c} \right)^3.\label{MC_fluc}
\end{eqnarray}
Therefore for a 3D harmonic trap the CN fluctuations, \eqref{CN_fluc}, are about 62\% larger in standard deviation than the MC ones,  as illustrated in Fig.~\ref{fig:graphical-abstract}, while the mean condensate fraction is unchanged. The variance ratio is 
\begin{equation}\label{SSS}
S =\frac{\delta_{MC}^2N_0}{\delta_{CN}^2N_0} = 0.39.
\end{equation}

The Maxwell Daemon ensemble gives a correct description of the Bose system below the condensation temperature by fixing the glitches of the textbook description of the Bose-Einstein condensation of the ideal gas. It is also a powerful tool, allowing to study systems characterised, in principle, by any single-particle spectrum, see~\cite{Weiss1997,Grossmann1997}. 

\subsection{Spectral classification}
\label{SPEC}

Remarkably general conclusions have been gleaned from an analysis based purely on the functional form of the energy spectrum in a  trap in $d$ dimensions, following~\cite{deGroot50}:
\eq{spec}{
\epsilon_{\vec{j}} = \hbar\sum_{l=1}^d \omega_l j_l^{\sigma},
}
where $\sigma=1$ for a harmonic trap with the energy scale $\hbar \omega _l$ and $\sigma=2$ for a uniform gas in a box where $\hbar \omega _l\equiv h/(2m L_l^2 )$, $m$ is mass of boson and $L_l$ the box size in the $l$ direction.
The condensate occupation for $d>\sigma$ is $\langle N_0\rangle \propto N[1-(T/T_c)^{d/\sigma}]$. However, perhaps the most wide-reaching conclusion of the spectral analysis is that there are several distinct trap/dimension regimes for the ultracold temperatures that lie in the range $\ve_1\ll T\lesssim T_c$~\cite{Weiss1997,Grossmann1997}:

(1) High-dimensional traps with $d>2\sigma>0$, in which condensate  fluctuations are normal, Gaussian and follow the Central Limit Theorem (CLT) in the thermodynamic limit, but grow faster than linearly with temperature:
\eq{highd}{
\fluct{N_0} \propto N \left(\frac{T}{T_c}\right)^{d/\sigma}
}
Here the heat capacity is discontinuous at the transition temperature and  $\delta N_{\rm ex}=\delta N_0\sim\sqrt{\langle N_{\rm ex}\rangle}$ follow the square root law but in the number of excited particles $N_{\rm ex}$ not the total.
In this regime MC fluctuations have the same scaling but 
a smaller prefactor than CN due to the energy constraint~\cite{Navez1997}.

(2) Lower dimensional traps with $\sigma<d<2\sigma$, in which condensate fluctuations are anomalously large, and non-Gaussian in the thermodynamic limit:
\eq{lowd}{
\fluct{N_0} \propto N^{2\sigma/d} \left(\frac{T}{T_c}\right)^2.
}
Notably in the region $\sigma<d<2\sigma$, a BEC already exists, but the fluctuations are anomalously non-Gaussian in the canonical ensemble, though still weaker than the $\fluct{N_0} \propto N^2$ found in the GCE. Heat capacities are continuous and $\delta N_{\rm ex}=\delta N_0$ does not follow a square root law. In fact $\delta N_{\rm ex}\sim\langle N_{\rm ex}\rangle^{\sigma/d}$.

(3) In the regime $d<\sigma$ no condensate exists in the thermodynamic limit, but the scaling of fluctuations can be found as
\eq{lowlowd}{
\fluct{N_0} \sim \sum_{j=1}^d \left(\frac{T}{\hbar\omega_j}\right)^2,
}
heat capacity is continuous and the fluctuations are far from a square root law $\delta N_{\rm ex}\sim\langle N_{\rm ex}\rangle$. 

 The boundary case of $d=2\sigma$ contains a logarithmic dependence~\cite{Grossmann1997}
\eq{log2n0}{
\fluct{N_0} \propto
N\left(\frac{T}{T_c}\right)^2
\log\left(\sqrt{N}\,\frac{T}{T_c}+\gamma+1\right),
}
with $\gamma\approx 0.5772$, Euler's constant.
Remarkably, for low dimensional traps $d<2\sigma$, not only do the same scaling exponents as above occur for both the CN and MC ensembles, but even the prefactors are the same~\cite{Grossmann1997}.
The Maxwell deamon combined with a saddle point approximation allows to derive from physical grounds the famous Hardy and Ramanujan formula for the asymptotics of the partition number.
More detail, such as CN prefactors, is given in~\cite{Weiss1997}.

\subsection{Historical survey }
\label{MISC_IDGAS}

To flesh out our coverage, we relate here a selection of  results on fluctuations in the ideal gas, and their historical development since the 1990's.
Some but not all of these results can be also found in the earlier topical reviews~\cite{Ziff1977, Kocharovsky2006, Yukalov04,Yukalov24}.

Early work is described in detail in the extensive review of~\cite{Ziff1977}, including fluctuation results from Refs.~\cite{Fierz56,deGroot50,Dingle49,Dingle52,Fraser51,Hauge70,Reif65}, and we will not cover the results from this period in detail. Of particular note are a variety of recurrence relations that were reported by~\cite{Landsberg61} (MC) and later~\cite{Borrmann93,Brosens96,Wilkens97,Weiss1997,Chase99}, as well as the work of~\cite{Brosens96} which studied finite-size but large systems up to $N=10^4$ numerically, and in particular found CN specific heats in 1D and 3D harmonic traps.  

However, an explosion of work on condensate fluctuations occurred in the period 1996-1998 after the first experimental realizations of Bose-Einstein condensation. 
The early work~\cite{Grossmann1996} studied the MC fluctuations in a 1D harmonically trapped gas using asymptotic formulas from number theory and found several now well known features such as linear growth of $\delta N_0$ with $T$ until a maximum just below $T_c$, localization of the $N_0$ distribution, as well as the great discrepancy with the GCE. 
Gajda \textit{et al.}~\cite{Gajda97} applied the saddle point method to the (at the time) more experimentally accessible 3D trapped ideal gas in the MC, obtaining an analytic estimate of the full distribution $\mathcal{P}(N_0)$ and numerically investigating fluctuations, confirming $\fluct{N_0} \propto N$ and the presence of the peak below $T_c$.

A large part of the flurry of new results was based on the incisive idea of Politzer~\cite{Politzer1996} to consider explicitly the counting statistics of the excited particles and derive condensate statistics as a consequence, as well as its later rigorous formulation by Navez~\cite{Navez1997} who introduced the concept of the Maxwell's Demon ensemble. 
In thie latter paper it was used to rigorously study the relationship between MC and CN fluctuations, finding \eqref{MC_fluc} in the 3D trap as well as the general relationships
\eq{dNMCCN0}{
\langle\delta^2 N_0\rangle_{MC} = \langle\delta^2 N_0\rangle_{CN}
-\frac{\langle\delta N_0\delta E\rangle_{CN}}{\langle\delta^2 E\rangle_{CN}}
}
and $\langle N_0\rangle_{CN}=\langle N_0\rangle_{MC}$
in the thermodynamic limit, as well as confirming qualitative agreement with finite-N condensates. Here $\langle N_{\rm ex}\rangle=\langle N_0\rangle$, so \eqn{dNMCCN0} equals \eqn{dNMCCN}. Higher terms of \eqn{dNMCCN0} were later determined in~\cite{Holthaus99}.

In parallel,~\cite{Wilkens97} showed the presence of the now very well known and characteristic fluctuations peak below $T_c$ in the CN  for harmonic and box traps, the independence of $\delta N_0$ from $N$ at low $T$, and the narrowing of the $\mathcal{P}(N_0)$ distribution. For a 1D harmonic trap they also demonstrated that the CN ideal gas system's Ginzburg-Landau functional shows behaviour characteristic of a weakly interacting Bose gas despite having no true interaction.
Any particle that leaves the condensate must show up in the excited states, a process that resembles as if a collision had occurred. Values of $\delta N_0$ for the MC were in turn found by~\cite{Grossmann97prl} after developing an appropriate numerical procedure. These showed the same characteristic independence of $N$ at low $T$ as in the CN (See Fig.~\ref{fig:Holthausprl}). However, their functional dependence on $T$ and $N$ were found qualitatively similar but lower than the CN ones.~\cite{Grossmann97arxiv} looked at MC fluctuations in harmonic traps and showed that their magnitude is essentially independent of $N$, indicating the dubious nature of a GC ensemble model.

\begin{figure}
    \centering
    \includegraphics[width=0.8\columnwidth]{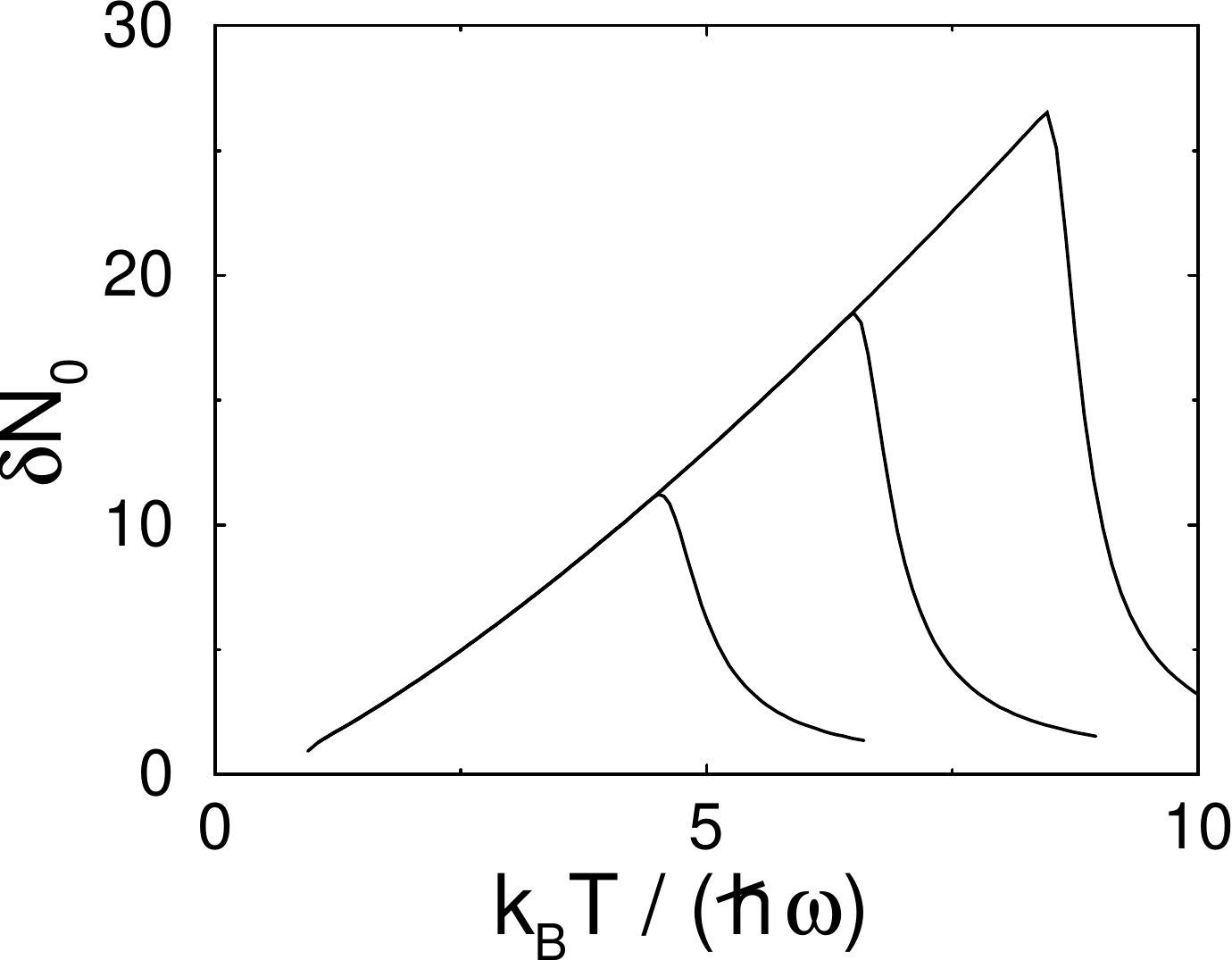}
    \caption{The change of microcanonical fluctuations $\delta_{MC}N_0$ with $N=200,500,1000$ (left to right) for an ideal gas in a 3d harmonic trap. Note the perfect overlap at low $T$.
    Reproduced from ~\cite{Grossmann97prl}
    \copyright American Physical Society, \href{https://dx.doi.org/10.1103/PhysRevLett.79.3557}{DOI}.
    }
    Used with permission.
    \label{fig:Holthausprl}
\end{figure}

The work \cite{Kocharovsky2006} introduced a so-called ``grand'' canonical approximation to CN quantities in which the mean condensate occupation
\eq{N0gen}{
\langle N_0\rangle = N - \sum_{j\neq0}\frac{1}{e^{\beta\varepsilon_j}-1},
}
is made to determine an effective fugacity $z_{N_0}=N_0/(N_0+1)$ consistent with \eqref{N0z} which then enters GC expressions for observables. This leads to
\eq{CNflucgen}{
\fluct{N_0} = \sum_i 
\frac{e^{\beta\varepsilon_i}/z_{N_0}}{\left[e^{\beta\varepsilon_i}/z_{N_0}-1\right]^2}.
}
The two concurrent works of Weiss \& Wilkens~\cite{Weiss1997} and Grossmann \& Holthaus~\cite{Grossmann1997} in a focus issue of the 1st volume of \textit{Optics Express} generalized and systematized fluctuation results to date, basing on the Politzer idea~\cite{Politzer1996} and the Maxwell Demon ensemble~\cite{Navez1997}. Notably, the essential dimensional classification based on \eqn{spec} was developed in full for the CN and MC~\cite{Grossmann1997} ensembles, as described in Sec.~\ref{SPEC}, and very general results for prefactors in different geometries were provided. Ref.~\cite{Weiss1997} also studied exact particle counting statistics and gave a lucid illustration of the fluctuations' dependence on $T$ and $N$. The work in Ref.~\cite{Grossmann1997} drew attention to the fact that the condensate fluctuations never depend on total $N$ but are instead tied to the excited population $\langle N_{\rm ex}\rangle$, following the ``expected'' square root law $\delta N_0\propto \sqrt{\langle N_{\rm ex}\rangle}$ for high dimensional traps ($d>2\sigma$) but growing faster than this like $\langle N_{\rm ex}\rangle$ for lower dimensional trap spectra. This work also
 detailed the relationship of fluctuations to heat capacity and connected the topic to questions in number theory.  
Further investigation of these avenues as applied to harmonic traps was carried out in both CN and MC ensembles by~\cite{HOLTHAUS1998198} using the poles of a Zeta function, with particular attention paid to anisotropic harmonic traps and the link to integer partitioning.

The saddle point method used in many of the above studies to describe systems in the thermodynamic limit requires modification and care in the presence of a condensate. 
\cite{Holthaus99} describes this in detail, and finds general results for the CN ensemble that indicate that the crossover across the critical temperature occurs for many observables in an error-function-like manner that becomes a step function as $N\to\infty$.

In a new development,~\cite{Scully99} introduced a master equation approach that has analogies to the quantum theory of a laser, as described in Sec.~\ref{MASTER}. This was soon developed further in~\cite{Kocharovsky2000c} which provides many detailed results in the ideal gas, and additionally benchmarked in~\cite{Holthaus01}. Follow on work~\cite{Scully06} then~\cite{Svidzinsky2006} (and more rigorously~\cite{Svidzinsky2010}) extended this to weakly interacting gases via a hybridization with the number-conserving Bogoliubov approach of~\cite{Kocharovsky2000}, obtaining well behaved properties also at temperatures of the order of, and across $T_c$.

Other relevant later work includes a detailed study of the GC vs CN, plus also partially MC, ensembles near the critical temperature~\cite{Tarasov15,Kocharovsky16} including consideration of universality classes,  building on earlier work~\cite{Kocharovsky10,Tarasov16}. The ensembles are found to disagree also in this region even in the thermodynamic limit. In related work, mesoscopic systems were found to exhibit non-Gaussian condensate statistics~\cite{Kocharovsky09}.
The ideal gas fluctuations were also re-analysed using a slightly different approach in~\cite{Xiong01,Xiong02} as background for analysis of weakly interacting cases. Fluctuations were briefly treated for the case of non-extensive Tsallis statistics in~\cite{Megias22}.
A recent work considered the CN ideal gas in a 3d box at fixed total momentum~\cite{Plyaschenik24}, finding condensate fragmentation at the lowest temperatures, and approximately Gaussian distributions for the occupations of condensed modes.

Yukalov, in a series of works~\cite{Yukalov04,Yukalov04b,Yukalov2005a,Yukalov05b}, has argued that systems with ``anomalous'' fluctuations $\fluct{N_0} \sim N^{a>1}$, e.g. ideal Bose gases with $2\sigma>d$ in the spectral classification, are unstable for thermodynamic reasons.
They then discussed in depth in~\cite{Yukalov05b} the argument that only stable descriptions of systems possessing $\fluct{N_0} \sim N^{a\le1}$ can fulfill the requirements of ``representative'' ensembles~\cite{Gibbs31,terHaar54,terHaar61}, and hence are indicated to submit to the principle of ensemble equivalence. A classification of this stability which is congruent with the spectral one of~\cite{Weiss1997,Grossmann1997} in Sec.~\ref{SPEC} was rederived in~\cite{Yukalov19}.
On the other hand the analysis of~\cite{Zhang06} pointed to the anomalous fluctuations being tied to a broken U(1) gauge symmetry, and argues that it does not imply the instability of the system. There a random phase approximation restores U(1) gauge symmetry, and makes anomalous fluctuations absent.

Finally, recent work has also begun to look into condensate fluctuations in polariton condensates \cite{Klaas2018, Bobrovska15, Alnatah2024Mar}. These are inherently open systems in which the total number of bosons, condensed or not, is not conserved, even on short timescales. Polariton number statistics in the steady (not equilibrium) state has been calculated and fluctuations analysed with the help of the Mandel $Q=\fluct{N_0}/\langle N_0\rangle -1$ parameter known from quantum optics~\cite{Zhang22,Wang22}.

\section{Modern frameworks for accounting fluctuations in bosonic systems\label{sec:modern-frameworks}}

In the previous section, we reviewed the definitions of statistical ensembles and highlighted the subtleties related to their inequivalence, particularly in the case of Bose-Einstein condensation. We also presented approximate asymptotic formulas for condensate fluctuations in the case of an ideal gas. It might seem that the matter of the ideal gas is thus settled. Unfortunately, it turns out that asymptotic formulas do not apply to the description of current experiments, even when the interactions between particles are so weak that the system can be considered an ideal gas. The problem lies in the fact that statistical quantities, such as condensate fluctuations, converge very slowly to their asymptotic values. As shown in~\cite{Christensen2020}, even for $N=100\,000$ atoms, the maximum fluctuations in a spherical trap differ from the asymptotic predictions of the fourth statistical ensemble by $20$\%. The discrepancy worsens for elongated traps (which are relevant in experiments). In an extreme case, in a 1D system, exact results~\cite{Weiss2002Aug} indicate that the asymptotic predictions regarding the probability distribution accurately reflect the true distribution only for $N$ on the order of $10^{10}$.

In practice, it is therefore necessary to compute statistical quantities for finite systems, which can usually be done only numerically. In this section, we present numerical techniques for generating statistical ensemble samples for a finite number of particles.

\subsection{Stochastic ensemble sampling}

The methods and the results concerning fluctuations of the Bose-Einstein condensate presented so far have two serious limitations. Firstly, they 
struggle to extend seamlessly to weakly interacting gases. Additionally, even for an ideal gas, numerical access to exact microcanonical fluctuations is restricted to a modest number of atoms, typically not more than 1000 in practice. 

Instead, a number of alternative stochastic approaches have been developed to represent the complete phase space of the multiatom Hilbert space by an ensemble of unbiased samples.
To do so one can 
generate numerous 
samples of configurations of the multiatom system, with their distribution constrained by a suitable control parameter. Among several approaches we focus here on those employing an occupation or classical field representation of atoms in modes as these are arguably the most flexible and most readily scalable to large but finite numbers $N\gg1000$.
In the following sections, we will outline three realizations of a Metropolis sampling approach:
the first, based on a classical fields approximation, 
the second, known as the Fock States Sampling method, which models the phase space using a suitable set of number states below, and a hybrid approach that keeps the good features of both in Sec.~\ref{HYB}.
We will also describe the master equation approach in Sec.~\ref{MASTER}, and briefly mention the method of ergodic dynamical evolution of classical fields in Sec.~\ref{ERG}.

\subsection{Metropolis sampling in the classical fields approximation}

 For equilibrium thermal states in 
 the canonical ensemble, one employs the Boltzmann factor for the distribution. To construct the canonical set efficiently, one can apply the Metropolis algorithm~\cite{Metropolis1953}, taking care of detailed balance conditions and thus defining a proper Markov chain that is subsequently sampled. Transition to the microcanonical ensemble then merely involves a straightforward 
post-selection 
from the 
canonical ensemble. 
For example, we can extract a subset of states with a well defined narrow energy range forming a representation of the microcanonical ensemble. Once these ensembles are established, calculating probability distributions for various variables becomes a direct averaging process over the samples.
\begin{figure}
    \centering
    \includegraphics[width=0.5\textwidth]{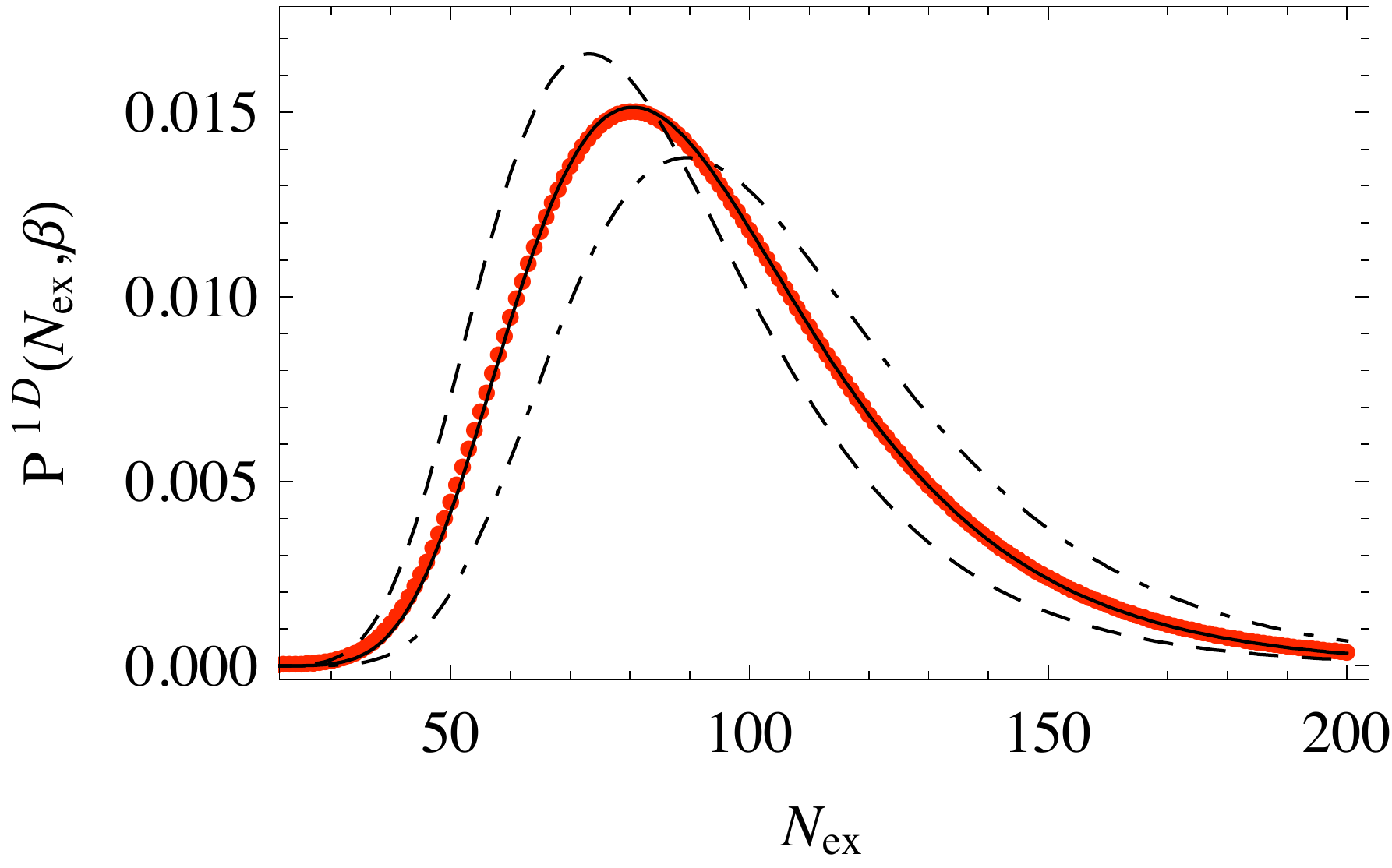}
    \caption{
    Probability distribution of having $N_{\rm ex}$ bosons outside the ground state of a 1D trap. Points represent the exact quantum distribution while lines are their classical field counterparts with $\hbar \omega j_{\rm max}=0.9\kB T$ (dotted-dashed line), $\hbar \omega j_{\rm max}=\kB T$ (solid line), and $\hbar \omega \beta j_{\rm max}=1.1\kB T$ (dashed line).  $N=1000$ and temperature $\hbar \omega \beta
    =0.04$.
    Reprinted (figure) with permission from Witkowska, Gajda, Rz\k a\.zewski~\cite{PhysRevA.79.033631}, \copyright  American Physical Society. Used with permissionof all Authors of~\cite{PhysRevA.79.033631}.}
        \label{fig:pnex}
\end{figure}

The quantum theory of multiparticle systems finds a
convenient formulation through the use of an atomic
quantum field in 2nd quantization denoted as $\oppsi$. 
We can expand
the atomic field in the single particle states with wavefunctions $\phi_j(\bm{r})$ and annihilation operators $\hat{a}_j$:
\begin{equation}
\oppsi = \sum_j \phi_j (\bm{r}) \,\hat{a}_j
\label{eq:oppsi}
\end{equation}
The classical field approximation, reviewed in detail in~\cite{Brewczyk2007Jan} and~\cite{Blakie08}, applies to bosons and involves replacing
operators $\hat{a}_j$ of highly occupied single-particle states in  \eqref{eq:oppsi} with
complex amplitudes as per $\hat{a}_j\to\alpha_j$,
while discarding other terms. This
results in replacing the entire field operator with a complex number
wave function 
\begin{equation}\label{cfPsi}
    \Psi(\bo{r}) = \sum_j \phi_j(\bo{r})\,\alpha_j.
\end{equation}
It corresponds to treating the wave aspects of the wavefunction analogously to classical electrodynamics for a photon field while discarding the ``particle-like'' effects, and becomes accurate as occupations become high enough.
Notably, in thermal equilibrium, the highly
occupied states correspond to those with the lowest
energies, which are of most interest, especially for condensate fluctuations. 

In order to generate a CN ensemble the probability (weight in the partition function) of any given classical field configuration $\vec{\alpha}=\{\alpha_j\}$ in the CN is
\begin{equation}
 P_j\bb{\alpha_j} = \frac{1}{Z_{\rm CFA}(N,T)}\exp\bb{-\frac{\sum_j\,\vej|\alpha_j|^2}{\kB T}}
\end{equation}
where $Z_{\rm CFA}(N,T)=\sum_{\{\alpha_j\}} P\bb{\{\alpha_j\}}$ is the classical
partition function and $\vej$ the energy of single-particle state $j$. The amplitudes $\alpha_j$ are subject to the
constraint:
\begin{equation}\label{mcnorm}
\sum_j |\alpha_j|^2 = N.
\end{equation}
A Metropolis sampling algorithm for $\vec{\alpha}$ is readily implemented as described in~\cite{Witkowska2010Mar}. Proposed updates take e.g. the form $\alpha_j\to\alpha_j+\delta_j$ with Gaussian complex noise $\delta_j$ whose amplitude is chosen for efficiency of the algorithm. One can update a single randomly chosen mode at a time or all together, correcting to preserve normalization \eqref{mcnorm}. 
The procedure is also easily generalized to the GCE~\cite{Pietraszewicz18a, Gawryluk2021May, Gawryluk2024May}. 

The classical fields approximation requires care, however, due to the appearance of a UV catastrophe in high energy modes. The set of modes should be terminated at single-particle energies of the order of $k_BT$ to avoid a preponderance of Rayleigh-Jeans distributed mode occupations ($\langle N_j\rangle\sim k_BT/\ve_j$) in the high energy tails since the correct distribution is the Bose-Einstein one. Terminating the single-particle states
at a cut-off can be 
a significant drawback as care becomes required to control the amount of influence the cutoff exerts~\cite{Blakie08,Cockburn11a}, especially as dimensionality grows~\cite{Brewczyk2007Jan}. For example, different energy cutoffs are optimal for different observables~\cite{Pietraszewicz15,Pietraszewicz18a,Pietraszewicz18b}, and one must balance the mode cutoff in position and momentum space~\cite{Bradley05}.

A benchmarking of the classical field representation in
a one-dimensional harmonic trap (frequency $\omega$) is illustrated in Fig.~\ref{fig:pnex}.
For this geometry the classical field result can be found in closed form
\cite{Bienias11a}:
\begin{equation}\label{cf-analytic-1dtrap}
    \mathcal{P}(\nex, T) = \frac{\xi^{\nex}}{1-\xi^{N}}\bb{\frac{1-\xi^{\nex}}{1-\xi^{N}}}^{j_{\rm max}-1}
\end{equation}
where $j_{\rm max}$ is the cut-off parameter (maximum value of $j$ included), and $\xi:=e^{-\frac{\hbar \,\omega}{\kB T}}$.
As shown in Fig.~\ref{fig:pnex}, the result \eqref{cf-analytic-1dtrap} coincides well with the exact distribution for the cut-off \begin{equation}\label{cutoff}
    E_{\rm max} = j_{\rm max}\hbar\omega = \kB T,
\end{equation}
but the match degrades as the cutoff value is moved away from the optimal one.

\subsection{Fock state sampling method\label{subsec:fssm}}
The fact that
the classical fields approximation 
neglects the corpuscular nature of atoms makes it
bear similarity to the pre-Planck
approach to black body radiation, leading to a Rayleigh-Jeans distribution of mean occupations at high energy $\vej\gtrsim k_BT$, and statistical
outcomes plagued by an ultraviolet catastrophe if $E_{\rm max}\gg k_BT$. 
Similarly, restoring the corpuscular character of atoms is able to cure
the ultraviolet divergence in our case. 
This crucial aspect is
a distinctive feature of the Fock States Sampling (FSS) method developed recently~\cite{Kruk23} which we will now outline.

While an exact theory would demand the parametrization of
the complete multiparticle Hilbert space (intractable for large systems), here one, opts for a model
based on a skeleton of Fock states:
\begin{equation}
    \ket{N_0,\,N_1,\, \ldots N_K},
    \label{eq:fock-KR}
\end{equation}
where $N_j$ are occupations of single-particle orbitals (modes). 
The Markov chain framework encompasses all distributions of $N$ atoms among the
orbitals. Numerics impose an energy cut-off $K$ in this context as well,
yet, unlike a classical field representation the construction converges with $K\to\infty$, rendering results cut-off
independent when the cut-off is suitably large.
Since states (\ref{eq:fock-KR}) are the eigenstates of the Hamiltonian of the noninteracting gas, the Boltzmann factor for a canonical ensemble reads:
$ \left( N_j\right) = \exp\left( -\beta \sum_j N_j \vej 
\right). 
$

\begin{figure}
    \centering
    \includegraphics[width=\linewidth]{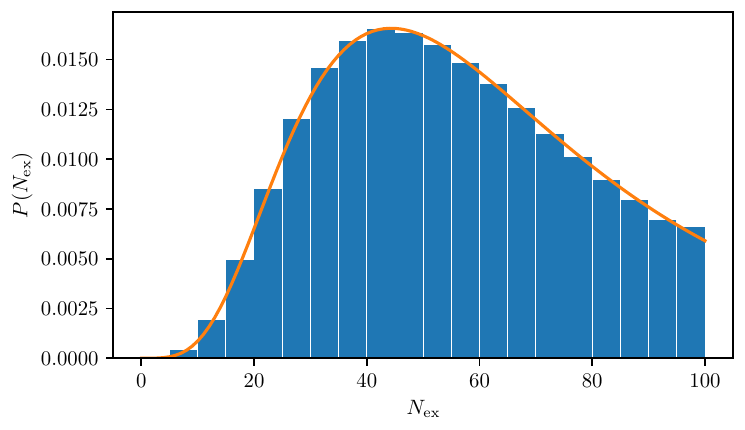}
    \caption{Distribution $\mathcal{P}(\nex)$ of the number of excited atoms  for $N=100$ bosons in a one-dimensional box with impenetrable walls for temperature $k_BT=110 \hbar^2/(2mL^2)$, where $L$ is the with of the box. Blue histogram -- FSS, solid orange line -- exact calculation using \eqref{eq:weiss-wilkens-recurrence}.}
    \label{fig:KR3}
\end{figure}

A Markov chain requires a choice of the Metropolis “dynamics” -- proposed states and acceptance criteria, which 
must satisfy several mathematical conditions. First requirement is the 
the principle of detailed balance, i.e. that the probability of a transition must be equal to the probability of its reverse. 
Secondly, the “dynamics” may not leave any state unreachable 
by the algorithm. Numerically, the most efficient 
algorithms are found to be based on the properties of bosonic dynamical processes. The probability of proposing a new orbital 
should be equal for every atom. Thus, it 
should be proportional to the current number of atoms in a given mode. The probability of acceptance 
in a given mode should be proportional to the number of atoms in the mode plus one, 
imitating the well-known Bose enhancement factor known from 
stimulated and spontaneous processes, dating back to the relation of A and B coefficients introduced by A. Einstein~\cite{Einstein1916}. Thus, the probability of proposing a jump from orbital $j$ to orbital $l$ reads:
\begin{equation}
    p_{j,\,l} \propto N_j \bb{N_l+1}
\end{equation}
The algorithm is completed by a standard acceptance criterion. We draw a random number $r$ from the interval $[0,1]$ and accept the jump if
$\exp(\beta\left[ E\bb{N_j}-E\bb{N_l}\right]) \geq r
$.

\begin{figure}
    \centering
    \includegraphics[width=\linewidth]{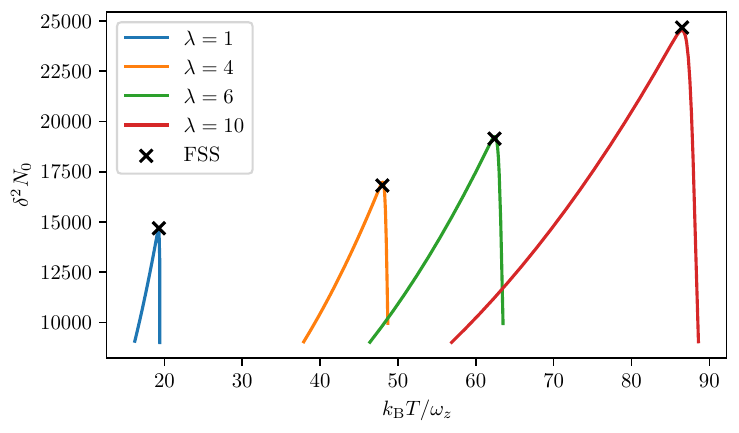}
    \caption{
        Fluctuations of BEC in a cloud of $N=10 000$ noninteracting atoms as a function of temperature in 3D traps with different aspect ratios $\lambda$. The solid lines are exact results obtained using \eqref{eq:weiss-wilkens-recurrence}. Results marked with black crosses were obtained via FSS method at the point of maximal fluctuations (most difficult to calculate). }
    \label{fig:KR4}
\end{figure}

The quality of results obtained with the FSS method is illustrated in Fig.~\ref{fig:KR3}, which  compares the FSS-produced histogram to the exact probability distribution of excited atoms confined in a one-dimensional box with impenetrable walls. Fig.~\ref{fig:KR4} shows the temperature dependence of $\delta N_0$ for $N=10000$ atoms in an elongated three-dimensional harmonic trap of various aspect ratios $\lambda=\omega_x/\omega_z=\omega_y/\omega_z$ as relevant for current experiment~\cite{Christensen2021}. 
The exact results (solid lines) are obtained with the help of the recurrence relations (\ref{eq:weiss-wilkens-recurrence}), while the FSS results are marked with the stars. The agreement is excellent.

\begin{figure}
    \centering
    \includegraphics[width=6cm]{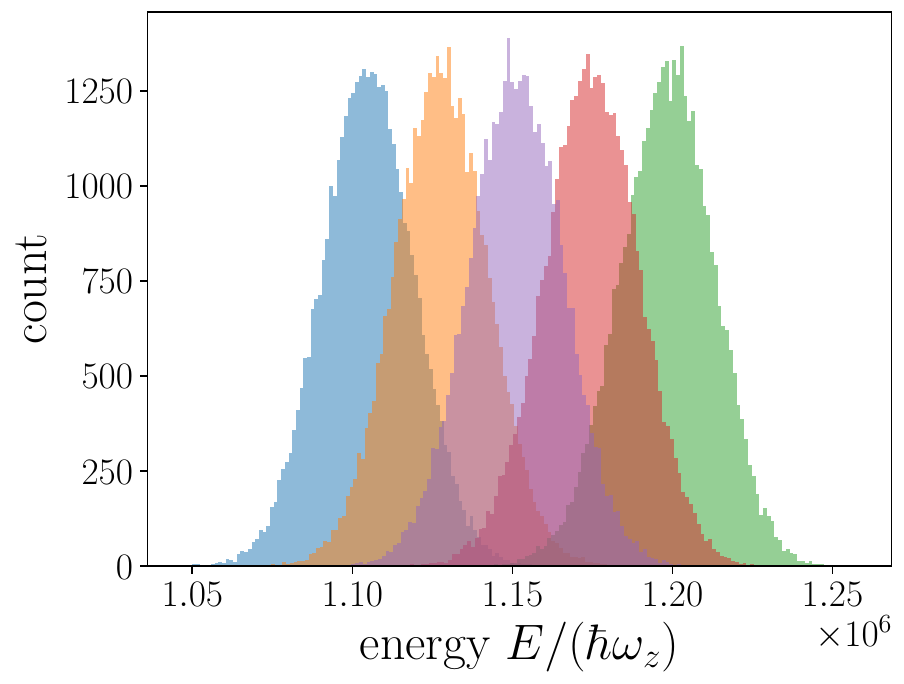}
    \caption{Histograms depicting the energy distributions of CN sets of states generated at different temperatures using the FSS method. The ensembles correspond to $N=10^4$  atoms confined in a 3D harmonic trap with an aspect ratio $\lambda=4$. The temperatures, shown from left (blue) to right (green), are  $k_B T/\hbar \omega_z = 47.5,47.75,48,48.25,48.5$.
    Figure adopted from \cite{Kruk23}. \href{https://creativecommons.org/licenses/by/4.0/}{CC BY 4.0}. 
    }
    \label{fig:KR5}
\end{figure}

\begin{figure}
    \centering
    \includegraphics[width=8cm]{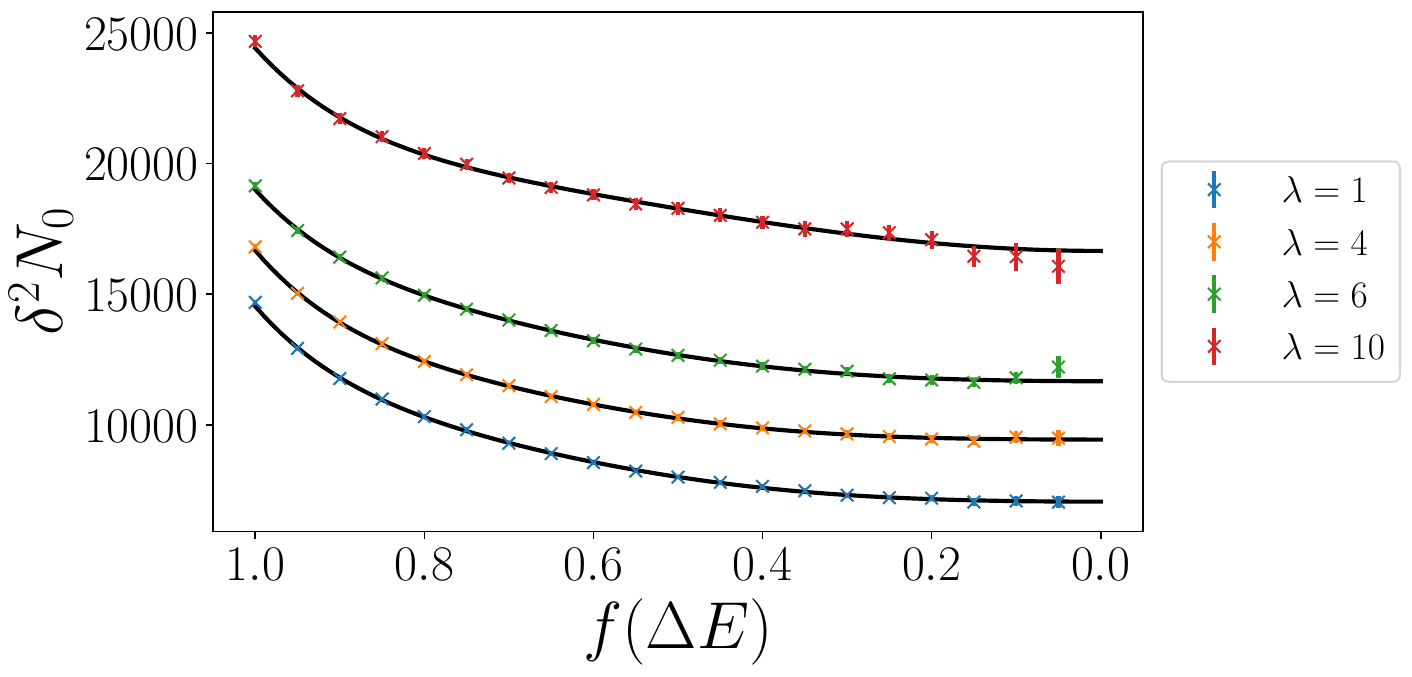}
    \caption{Dependence of condensate fluctuations on the energy window $\Delta E$ for $N=10000$ atoms and different aspect ratios $\lambda$. $f(\Delta E)$ is the fraction of microstates that are within the energy window $\Delta E$. Presented here is data obtained via postselection of FSS-generated states -- crosses with uncertainties, and fitted high degree polynomials -- solid black lines -- to best estimate the MC limit. 
} 
    \label{fig:KR6}
\end{figure}

\begin{figure}
    \centering
    \includegraphics[width=6cm]{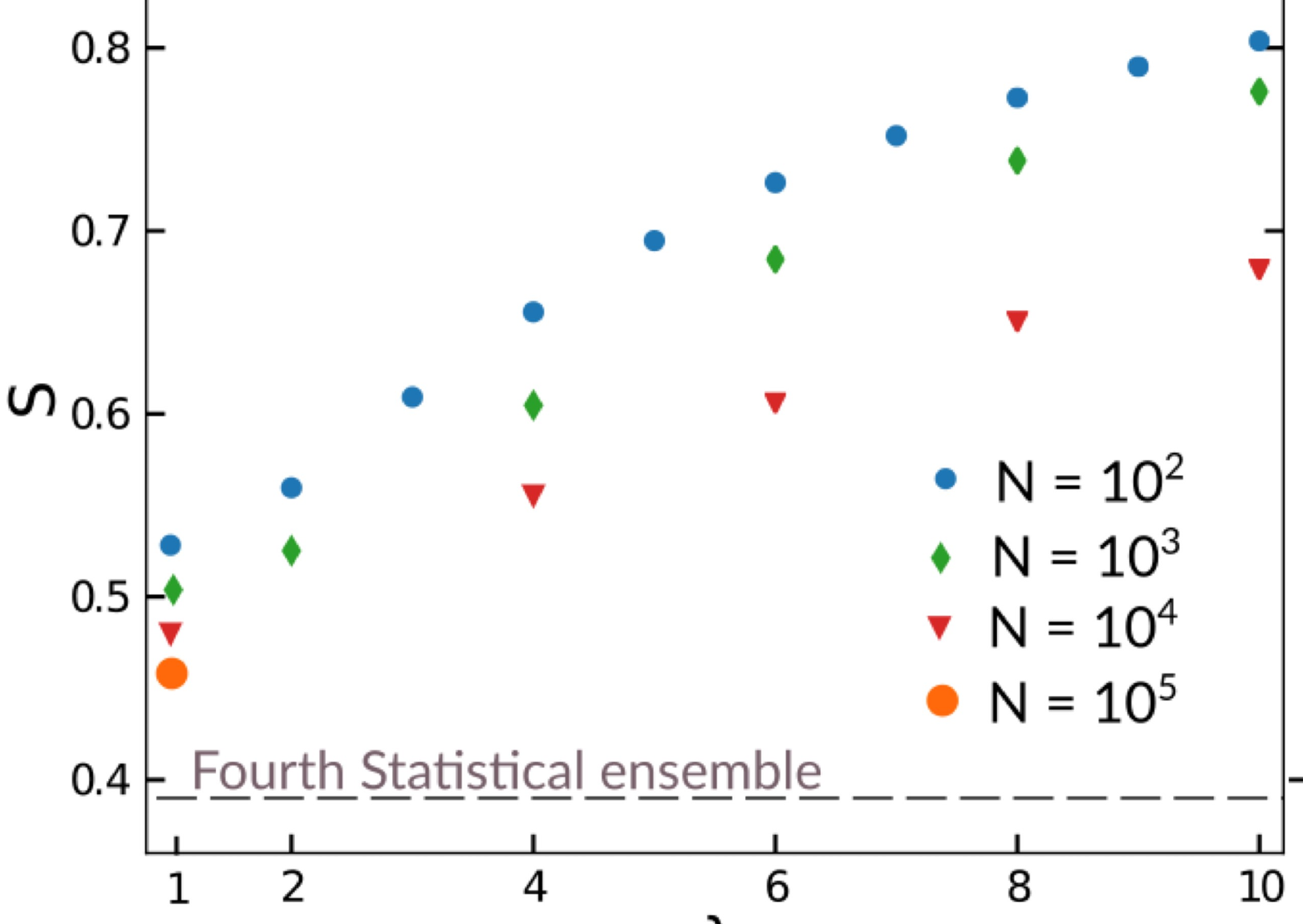}
    \caption{
 Ratio, $S$, between the calculated peak variance of the noninteracting BEC in the microcanonical and canonical ensembles as
a function of the trap aspect ratio $\lambda$ for $N = 10^2$ (blue points), $N = 10^3$ (green diamonds), $N = 10^4$ (red triangles) and $N = 10^5$
(purple square) atoms. The black dashed line corresponds to the asymptotic value estimated within the Maxwell’s demon ensemble (\ref{SSS}). 
The three open symbols correspond to experimental points, presented in \cite{Christensen2021}, with atom numbers $N = 8.8 \times 10^4 -- 10.4 \times 10^4$. 
Figure reproduced from  \cite{Christensen2021}. (\href{https://creativecommons.org/licenses/by/4.0/}{CC BY 4.0}).\label{fig:KR7}}
\end{figure}

The above results also offer a practical avenue for deriving microcanonical fluctuations in systems with a large number of atoms, where the computational complexity of employing recurrence relations becomes prohibitive. In 
Fig.~\ref{fig:KR5}, 
the energy probability distributions at different temperatures in the CN ensemble are shown. Through a post-selection process 
one can selectively discard system copies lying outside a narrowing energy interval 
$E\in [\wb{E}-\Delta_E/2, \wb{E}+\Delta_E/2]$
 around the distribution’s central energy $\wb{E}$. Although this decreases the number of retained copies, impacting statistical robustness, one can mitigate this effect by extrapolating results from larger intervals to estimate a well-defined fixed energy outcome. 
How this plays out can be seen in Fig.~\ref{fig:KR6}, where one also observes the continuum of ensembles between the CN and MC ensembles. 
lier or later chapters.}

Fig.~\ref{fig:KR7} presents an experimentally important quantity -- $S=\delta_{MC}^2N_0 / \delta_{CN}^2N_0$, the ratio of maximal microcanonical fluctuations to canonical fluctuations. It is shown for various aspect ratios and increasing total numbers of atoms in a harmonic trap. Notably, even for a spherical trap with up to $100 000$ atoms, the calculated $S$ exceeds significantly the $N\to\infty$ asymptotic value of $0.39$, \eqref{SSS}, as elucidated in~\cite{Christensen2021}.

\subsection{Dynamical ergodic approach and related methods}
\label{ERG}
A conceptually different, dynamical, approach  to generate ensembles is also often used using  evolution of classical field samples. For example, evolving a random field sample with total energy $E$ and $N$ particles in time via a Gross-Pitaevskii equation for the field (\ref{cfPsi}), 
\begin{equation}\label{GPE}
i\frac{d\Psi(\bo{r})}{dt} = \left[-\frac{\nabla^2}{2}+V(\bo{r})+g|\Psi(\bo{r})|^2\right]\Psi(x)
\end{equation}
with $V(\bo{r})$ the trapping potential and $g$ the inter-particle interaction.
Taking configurations at many consecutive times separated by an interval $\delta t$ allows one to obtain microcanonical (MC) ensemble samples~\cite{Goral01,Davis01a,Brewczyk07,Blakie08} relying on an assumption of ergodicity. The conserved number of particles is 
\eq{Ngp}{
N = \int d^d\bo{r} |\Psi(\bo{r})|^2,
}
so we can see that a run will give at most an MC ensemble. Evolving a stochastic Gross-Pitaevskii equation in contact with a bath at temperature $T$ and chemical potential $\mu$ through ``growth'' terms will generate GC samples~\cite{Blakie08,Gardiner03}, whereas having contact with the bath via only ``scattering'' terms will generate CN samples~\cite{Rooney12}; A GC-type stochastic Gross-Pitaevskii equation modified by the addition of an appropriate restoring mechanism on $N$ (sometimes called a nonlinear chemical potential) can produce ensembles lying between CN and GC having a variance of $N$ controlled by the user, for example to match actual variances produced by the preparation protocol in an experiment~\cite{Pietraszewicz2017}.   

Under the right conditions it is also possible to go beyond classical field samples to a set of configurations that includes the effects of quantum fluctuations. For example when $N_{\rm ex}\ll N$, Wigner distribution samples of a Bogoliubov description can be obtained~\cite{Sinatra00,FINESS-Book-Ruostekoski,Martin10b}, or a non-perturbative distribution of positive-P samples can be obtained via imaginary time evolution~\cite{Drummond04,Deuar09}, or under normal time dissipative evolution~\cite{Deuar21}. One can also generate number-phase Wigner samples~\cite{Heller09,Heller13}, and other varied schemes have also been proposed.

All the dynamical approaches rely on the presence of interactions to invoke ergodicty, unlike the Metropolis sampling and master equation formulations (below) that do not. However, the advantage of the ergodic approach for interacting systems is that it makes no hard-wired presuppositions about the one-particle basis states.

\subsection{Master equation formulation} 
\label{MASTER}

The master equation approach developed by Scully, Kocharovsky, and collaborators~\cite{Scully99,Kocharovsky2000c} for finite $N$ systems can provide a great deal of information about condensate and excited state fluctuations in a wide range of geometries, as well as analytic and numerical results for distribution functions or higher condensate moments. It
bears similarity to the quantum model of the laser, and also to a class of stochastic Gross-Pitaevskii equation models for interacting gases that includes only ``scattering'' and not ``growth'' processes~\cite{Rooney12}. The approach has been reviewed in
 detail in~\cite{Kocharovsky2006}. Here we recapitulate the main essence of the model and salient results on condensate fluctuations.

The model starts with a trapped ideal gas system with boson modes $\opr{a}_j$ ($j\ge0$) of energy $\ve_j$ (occupations $\hat{N}_j=\dagopr{a}_j\opr{a}_j$), in contact with a thermal reservoir of harmonic oscillator modes $\opr{r}_{\nu}$ with a dense energy spectrum having Bose-Einstein distributed mode occupations 
\eq{etanu}{
\eta_{\nu} = \left[e^{\hbar\omega_{\nu}/k_BT}-1\right]^{-1}.
}
The reservoir-system interaction Hamiltonian in the interaction picture with coupling strengths $g_{\nu,kl}$ is given by
\eq{coup}{
\opr{V} = \sum_{\nu}\sum_{l\ge0, k>l}
g_{\nu,kl}\ \dagopr{r}_{\nu} \opr{a}_k\dagopr{a}_l e^{-i(\hbar\omega_{\nu}-\ve_k+\ve_l)t} + {\rm h.c.},
}
which preserves particle number in the system, but not energy -- thereby providing a canonical ensemble. 
System modes $k,l$ scatter off reservoir quanta which are annihilated or created.
The total atom number is $N=N_0+\sum_{j>0}N_j$.
Assuming a large reservoir with a smooth dense energy spectrum, unchanged by the contact, and some auxiliary assumptions~\cite{Kocharovsky2000c}, one obtains the master equation for the reduced density operator of the system:
%
I will use jump operator, to stress that this is typical Lindblad form). Then equation will be
\begin{eqnarray}  
\nonumber\frac{d\opr{\rho}}{dt} &=& \frac{\kappa}{2}\sum_{k>l}
(\eta_{kl}+1)\left[
2\hat{C}_{kl} \opr{\rho} \hat{C}^{\dagger}_{kl}-\hat{C}_{kl}^{\dagger}\hat{C}_{kl}\opr{\rho} 
-\opr{\rho}\hat{C}_{kl}^{\dagger}\hat{C}_{kl}
\right] \\
\nonumber
& &+\frac{\kappa}{2}\sum_{k>l}\eta_{kl} \left[
2\hat{C}^{\dagger}_{kl}\opr{\rho}\hat{C}_{kl} - \hat{C}^{\dagger}_{kl}\hat{C}_{kl}\opr{\rho} 
- \opr{\rho}\hat{C}^{\dagger}_{kl}\hat{C}_{kl} 
\right],
\end{eqnarray}
where $\hat{C}_{kl}\equiv \dagopr{a}_l\opr{a}_k$ is a jump operator.

%
Here the shorthand 
\eq{etakl}{
\eta_{kl}= \left[e^{\hbar(\ve_k-\ve_l)/k_BT}-1\right]^{-1}
}
is used to indicate the reservoir occupation at an energy corresponding to the energy difference between levels $k$ and $l$ in the trap. Next, one assumes that the mean excited state occupations $\langle \hat{N}_{j>0}\rangle_{N_0}$ for a given condensate occupation $N_0$ are thermally distributed with temperature $T$ for each $N_0$ value, i.e. $\propto {\rm exp}[-\sum_{j>0}\ve_jN_j/k_BT]$.

It yields an equation of motion for the probabilities of condensate occupations:
\begin{eqnarray}
\frac{dP_{N_0}}{dt} &=& -\kappa\left\{ K_{N_0}(N_0+1)P_{N_0} - K_{N_0-1}N_0 P_{N_0-1}\right.\nonu\\  
& &\left.+H_{N_0}N_0 P_{N_0}-H_{N_0+1}(N_0+1)P_{N_0+1}\right\},
\label{dpn0dt}
\end{eqnarray}
where $K_{N_0} = \sum_{l>0}(\eta_l+1)\langle N_{l}\rangle_{N_0}$ and 
$H_{N_0} = \sum_{l>0}\eta_l(\langle N_{l}\rangle_{N_0}+1)$ are cooling and heating rates, respectively. The sums are in terms of Bose-Einstein occupation factors \eqn{etanu} corresponding to energies $\omega_k=\ve_k$.  This yields the quite general equilibrium distribution
\eq{Pnoeq}{
P_{N_0} = P_0 \prod_{j=1}^{N_0} \frac{K_{j-1}}{H_j}.
}

Closed-form results were obtained for two salient temperature regimes. 
At \underline{very low temperatures} such that $\eta_j\ll1$ and $\langle \hat{N}_j\rangle_{N_0}\ll1$, one finds the condensate distribution to be governed by only one essential parameter 
\eq{mcH}{
\mc{H} = \sum_{k>0} \eta_k = \sum_{k>0}\frac{1}{e^{\ve_k/T}-1} =\langle N_{\rm ex}\rangle_{CN}
}
corresponding to the mean number of excited quanta as a function of $T$. 
This holds for $T\ll\ve_1$ 
in general, and for a much wider temperature range in a harmonic trap. 
One has 
\eq{Pnocold}{
P_{N_0} = \frac{1}{{\cal Z}}\,\frac{\mc{H}^{N-N_0}}{(N-N_0)!}
}
with the partition function ${\cal Z}$ providing the normalization $\sum_{N_0}P_{N_0}=1$. This leads to
\eq{dn0cold}{
\fluct{N_0} = \mc{H}\left(1-\frac{e^{-\mc{H}}(\langle N_0\rangle+1)\mc{H}^N}{\Gamma(N+1,\mc{H})}\right)
}   
with $\Gamma(\alpha,x)=\int_x^{\infty}t^{\alpha-1}e^{-t}dt$ where $\langle N_0\rangle= N-\mc{H}\left[1+e^{-\mc{H}}\mc{H}^N/\Gamma(N+1,\mc{H})\right]$.

In more general \underline{higher temperatures} a ``quasi-thermal'' distribution is assumed for 
\eq{quasith}{
\langle N_{j>0}\rangle_{N_0} = \left(\frac{N-N_0}{\mc{H}}\right)\,\eta_j,
}
which differs from standard thermal mode occupations only by the prefactor that makes it satisfy $N=N_0+\sum_{j>0}\langle N_j\rangle_{N_0}$. Now one obtains heating and cooling coefficients
$K_{N_0} = (N-N_0)(1+\wb{\eta})$ and $H_{N_0}=\mc{H}+(N-N_0)\wb{\eta}$
and the whole distribution is governed by two parameters: $\mc{H}$ and the cross-correlation 
\eq{etadef}{\wb{\eta}= \frac{1}{N-N_0}\sum_{j>0}\eta_j\langle N_j\rangle_{N_0} = \frac{1}{\mc{H}}\sum_{j>0}\eta_j^2.
}
The probability distribution is binomial-related:
\begin{equation*}
P_{N_0}\ = \frac{1}{\cal Z}
\binom{N-N_0-1+\mathcal{H}/\wb{\eta}}{N-N_0}
\left(\frac{\wb{\eta}}{1+\wb{\eta}}\right)^{N-N_0}.    
\end{equation*}
The fluctuations are 
\begin{eqnarray}
    \fluct{N_0}2 &=& (1+\wb{\eta})\mc{H} -P_0(\wb{\eta} N+\mc{H})(N-\mc{H}+\wb{\eta}+1)\nonu\\&&-P_0^2(\wb{\eta} N+\mc{H})^2,
    \label{dn0hot}
\end{eqnarray}
and the mean $\langle N_0\rangle = N-\mathcal{H}+P_0(N\wb{\eta}+\mathcal{H})$
where 
\eq{P0result}{
P_0=\frac{1}{\cal{Z}}\left(\frac{(N+\mc{H}/\wb{\eta}-1)!}{N!(\mc{H}/\wb{\eta}-1)!}\right)\ \left(\frac{\wb{\eta}}{1+\wb{\eta}}\right)^N
}
is the probability of having zero ground state atoms. In the thermodynamic limit $P_0\to 0$, leaving us with just $\fluct{N_0} = (1+\wb{\eta})\mc{H}$ known previously~\cite{Ziff1977}.
Approximate expressions for $\mathcal{H}, \wb{\eta}$ in traps of various shape and dimensionality, in and not in the thermodynamic limit, are given in~\cite{Kocharovsky2006}.

The formalism of the master equation has been also used to study the statistical properties of ultracold gases during different cooling techniques see for instance ~\cite{Cirac1994} that showed that the mean occupations in the stationary state of a laser-cooled ideal gas will resemble Bose-Einstein statistics.
\section{Role of interactions on particle number statistics \label{INT}}
We shift now from the secure domain of the ideal gas to the more challenging realm of fluctuations in a weakly interacting gas.
We stress that there are no exact results in this case.
We will start by recalling the main 
results
of the Bogoliubov approximation for the weakly interacting gases with repulsive contact interaction strength $g$, and follow by presenting more advanced modern approaches.

As background, one needs to keep in mind several global aspects of the system that change when interactions appear. One is that in a trapped system the shape, in particular the width, of the condensate mode can change drastically. At temperatures $T\ll T_c$ it is well described by the stationary solution of the GPE \eqref{GPE}. Given sufficient central density $n_0$ such that $gn_0\gg\hbar\omega$, the $T=0$ condensate is quite well approximated by the Thomas-Fermi profile
\eq{TF}{
\Psi(\bo{r}) = \sqrt{\frac{\mu-V(\bo{r})}{g}} \to \sqrt{\frac{\mu}{g}}\left[1-\sum_{d=1}^D \left(\frac{x_d}{R_d}\right)^2\right],
}
with Thomas-Fermi radii $R_d=(1/\hbar\omega_d)\sqrt{2n_0/m}$, and assumes negligible kinetic contribution. 
This shape and width change naturally strongly affects quantum fluctuations, including fluctuations of the lowest energy condensate mode. It is also crucial for the shift of $T_c$ in trap. 

A second aspect to keep in mind during the whole discussion to follow is that in lower dimensions a sufficiently large cloud will become a quasicondensate in which phase coherence is shorter than the cloud extent, and several low energy modes are highly occupied.  

{In the previous section, we reviewed several methods for calculating fluctuations in the number of condensed noninteracting bosons. We presented methods and results in different statistical ensembles and  for various trapping potentials and different dimensionalities. Most of
these results were exact. 

Moving to weakly interacting gases in this section, 
brings us closer to realistic experimental conditions. Moreover, the statistics of the ideal gas would have no physical meaning without the weak collisions necessary for achieving thermal equilibrium. 
However, by entering the realm of weakly interacting gases, we move beyond exact results. To the best of our knowledge, no exact condensate fluctuations are known in closed form for any model of large systems with interactions.}

\subsection{Bogolubov approximation}

At very low temperatures, well below the critical temperature, we can use the Bogoliubov approximation, whose use for condensate fluctuations in ultracold atoms was initiated by Giorgini, Pitaevskii, and Stringari in~\cite{Giorgini1998}. In this approximation, the ultracold gas is divided into a stable condensate 
containing most of the atoms and a minority of uncondensed atoms represented by Bogoliubov quasiparticles. Thus, in the symmetry breaking Bogoliubov variant that we will use here, the atomic field operator is represented in the following form:
\begin{equation}\label{bog_sym_break}
    \hat{\Psi}(\bm{r}) = \sqrt{N}\,\varphi_0(\bm{r}) + \hat{\delta} (\bm{r}) 
\end{equation}
where $\varphi_0(\bm{r})$ stands for a condensate wave-function obtained via the stationary solution of the Gross-Pitaevskii equation and the quantum part $\hat{\delta}$ 
represents together the thermal and quantum depleted atoms that do not occupy the condensate mode. The fundamental assumptions then made are that terms $\mathcal{O}(\hat{\delta}^3)$ and higher in the Hamiltonian can be discarded because the overall condensate depletion is small ($T\ll T_c$, ). Further, in a stationary, equilibrium state terms $\mathcal{O}(\hat{\delta})$ turn out to be zero. The quantum part is  conveniently diagonalized by a combination of the creation and annihilation operators of the Bogoliubov quasiparticles in modes labelled $l$: 
\begin{equation}
    \hat{\delta} (\bm{r}) = \sum_l u_l(\bm{r})\hat{b}_l + v(\bm{r})\hat{b}_l^{\dagger},
\end{equation}
where the $u_l$ and $v_l$ functions are obtained by solving the
Bogoliubov-de Gennes equations which read
\begin{eqnarray}
    \nonumber\left[ \hat{K} + V(\bo{r}) +  2 N  g |\varphi_0 |^2 - \mu\right]u_l + Ng \,\varphi_0^2
    v_l &=& \hbar \omega_l u_l,\\
    \nonumber\left[ \hat{K} + V(\bo{r}) +  2 N  g |\varphi_0|^2 - \mu\right]v_l + Ng \,\varphi_0^2 
    u_l&=& -\hbar \omega_l v_l\\
     &~& \label{eq:BdG}
\end{eqnarray}
where $\hat{K}=-\frac{\hbar^2 \Delta}{2 m}$ is the kinetic energy, 
$\mu$ is the chemical potential, and $\hbar\omega_l$ the quasiparticle energies above the ground state. Notably, these excitations follow a spectrum that is phonon-like for $\hbar\omega_l\propto (1/\lambda_l)$ when $\hbar\omega\lesssim\mu$ and $\lambda_l$ is a characteristic wavelength, and become particle like with $\hbar\omega_l\approx \hbar k_l^2/(2m)$ at high energies. The excitations and fluctuations have different properties in the two regimes.

To impose the bosonic commutation relations $[\hat{b}_l, \hat{b}_j]=1$, one uses the constraint for $u_l$ and $v_l$ functions
\begin{equation}
    \int |u_l (\bm{r})|^2 d^Dr - \int |v_l (\bm{r})|^2 d^Dr = 1.
\end{equation}
The eigenvalue equations (\ref{eq:BdG}) are solved numerically, except for the uniform case which has plane wave solutions and $\omega=(k/2)\sqrt{k^2+4gn}$. The procedure is simplified if the problem has a symmetry. For instance for a spherically symmetric case the $u_l$ and $v_l$  functions are proportional to spherical harmonics and the diagonalization is reduced to a single, radial variable.

The excitations behave as independent quantum oscillators, and in the canonical ensemble, each oscillator is in thermal equilibrium. Then, we readily calculate the variance in the number of thermal atoms. 
\begin{eqnarray}
    \fluct{N_0} &=& \fluct{\nex} = \meanv{\nex^2} - \meanv{\nex}^2\\
    \nex&=&\int\langle\hat{\delta}^{\dagger}(\bm{r})\hat{\delta}(\bm{r})\rangle d^Dr = \sum_l \langle\opr{b}^{\dagger}_l\opr{b}_l+1\rangle\int d^D\bo{r} |v_l(\bo{r})|^2.\nonumber
\end{eqnarray}

Since the symmetry breaking variant of the Bogoliubov approximation (\ref{bog_sym_break}) violates particle number conservation, as the temperature increases, the number of uncondensed atoms rises while the condensate remains undepleted. Thus, and also because of the removal of high order terms, this approximation yields reliable results only at very low temperatures when $N_{\rm ex}\ll N$.  We illustrate the relation between the Bogoliubov approximation and the exact result on a simple case of the noninteracting gas confined in a three dimensional one dimensional harmonic trap (see orange line in Fig.~\ref{fig:hybrid_harmonic_3d}).

The pioneering Bogoliubov analysis of~\cite{Giorgini1998} for the interacting 3D Bose gas in a box found that 
\mbox{$\fluct{N_0} = 2\sqrt{\pi}(a N_0)^3V$} even at $T=0$, solely due to interactions (the so-called quantum fluctuations), where $g=4\pi a\hbar^2/m$. At temperature, 
\eq{gbox3d}{
\fluct{N_0} \approx 0.105\left(\frac{mT}{\hbar^2}\right)^2V^{4/3} + \frac{k_B T m^2c V}{3\pi^2\hbar^3} \log\left(\frac{Vm^3c^3}{\hbar^3}\right).
}
when $T\ll T_c$, and $c=\sqrt{g N_0/mV}$ is the speed of sound. The first term is half of the value in the ideal gas with periodic boundary conditions, and does not depend on the interaction $a$. The factor $2$ between interacting and non-interacting cases was attributed to the choice of statistical ensemble in \cite{Giorgini1998} or, the order of limits when going to thermodynamic limit and $g\to 0$ \cite{Castin2002Jul, Castin2001May}.
Notably, both results shows super-Poissonian fluctuatations $\fluct{N_0} \approx V^{4/3} \propto N^{4/3}$. 
In a 3D harmonic trap on the other hand,
\eq{gtrap3d}{
\fluct{N_0} \propto \left(\frac{T}{T_c}\right)^2\left(\frac{ma^2T_c}{\hbar^2}\right)^{2/5} \,N^{4/3}.
}
exhibiting also a $4/3$ power scaling but interestingly with a prefactor dependent on interaction, and no corresponding ideal gas term.
Either way, scaling of fluctuations with a power greater than one in the size $N$ or $V$ indicates anomalously large fluctuations that according to the analysis of~\cite{Yukalov05b} would be expected to be unstable and not obey ensemble equivalence~\cite{Yukalov05b}. 

The subsequent work of Kocharovsky \textit{et al} used the number-conserving Bogoliubov formalism of Girardeau and Arnowitt~\cite{Girardeau59} to obtain fluctuations and higher moments of the condensate occupation~\cite{Kocharovsky2000,Kocharovsky2000a}, including box and arbitrary potential traps. In particular, it was shown that the condensate statistics are never Gaussian, and that interactions suppress their magnitude, presumably for energetic reasons.

Other work  studied 2D and quasi-2D trapped gases~\cite{Xiong02b,Xiong03} finding $\fluct{N_0}\propto N^2$ plus extra terms $\mathcal{O}(N^{3/2})$, $\mathcal{O}(N)$ and arriving at the rather strange looking powers $\fluct{N_0}\propto N^{22/15}$ for a 3D trapped gas~\cite{Liu03} and  $\fluct{N_0}\propto N^{10/3}/(\ln N)^2$ for the 1D trapped case~\cite{Liu03b}. Some very simplified GCE models such as a single-mode condensate have also been studied~\cite{Cherny05}.

 It has also been found \cite{Idziaszek2005} that the difference between MC and CN fluctuations scales normally in $N$, so they become equivalent in the thermodynamic limit (but anomalous). This contrasts heavily with the trapped ideal gas for which MC and CN fluctuations stay different in the thermodynamic limit.

\subsection{Anomalous fluctuations controversy}

However, in contrast o all the above, the early work of Idziaszek \textit{et al.}~\cite{Idziaszek1999} found normal fluctuations $\fluct{N_0}\sim N$ and later~\cite{Yukalov04b,Yukalov2005a} studied a symmetry-broken Bogoliubov description of the weakly interacting gas and investigated its thermodynamic considerations. They argued that fluctuations are normal $\fluct{N_0}\le \propto N$ in both CN and GCE, including arbitrarily nonuniform~\cite{Yukalov09}, or mesoscopic~\cite{Yukalov19} cases. Hence they would be expected to be representative ensembles and obey ensemble equivalence~\cite{Yukalov05b}. 

This issue of whether the fluctuations of the weakly interacting gas as described by Bogoliubov are anomalous or not has been  the source of a controversy with many studies indicating anomalous fluctuations as above, 
but others finding normal Gaussian fluctuations $\fluct{N_0} \propto N$. 
On the anomalous side: Ref.~\cite{Giorgini1998} (extended to stronger interactions by~\cite{Meier99}), then~\cite{Kocharovsky2000,Kocharovsky2000a}, as well as a later CN analysis by~\cite{Xiong01} (box)~\cite{Xiong02} (harmonic trap): 
On the normal side~\cite{Idziaszek1999}, but also later work by~\cite{Yukalov04b,Yukalov2005a,Yukalov09,Yukalov19}.

The matter was reassessed in the low temperature regime by Idziaszek in~\cite{Idziaszek2005}, recovering the anomalous scaling predictions in the particle-number-conserving and traditional, symmetry breaking nonconserving theory, and in a wide range of cases: both in MC and CN ensembles, in uniform and harmonically trapped gases in 3D. The review~\cite{Kocharovsky2006} has also argued for the presence of anomalous fluctuations. 

A number of explanations for the two kinds of results for condensate fluctuations have been proposed.
Ref.~\cite{Xiong02} has pointed to the difference arising from whether a perturbation theory approach is used (giving normal condensate fluctuations due to thermal atoms in single-particle excited orbitals) or a full Bogoliubov approach that describes primarily fluctuations due to collective excitations of low energy modes (anomalous condensate fluctuations). This interpretation is also confirmed by~\cite{Idziaszek2005}. 
Nevertheless, Ref.~\cite{Yukalov24}, using a different analysis, argues that anomalous fluctuations are a result of inconsistent inclusion of higher order terms in the Bogoliubov calculations, and should not be present when $U(1)$ symmetry is broken. The earlier work of~\cite{Zhang06} pointed to anomalous fluctuations being tied to a broken $U(1)$ gauge symmetry, but also argued that appropriate use of the random phase approximation of~\cite{Minguzzi97} restores normal fluctuations. 
A further angle on the problem was provided by \cite{Idziaszek_2009} using a Bogoliubov-Popov approach which saw a crossover from anomalous to normal fluctuations for sufficiently large systems, though this same work also predicted unphysically large fluctuations near $T_c$.

Overall it appears that the controversy regarding anomalous or normal condensate fluctuations is convolved with the issue of whether a thermodynamic limit or an explicitly mesoscopic-size system is under consideration, but there are yet be further aspects of the matter to understand. 

\subsection{Beyond Bogoliubov}

Study of the statistics of interacting bosons beyond the regime of applicability of the Bogoliubov approximation is notoriously difficult. The exact result can be  obtained in the thermodynamic limit of the 1D box, modelled by the Lieb-Liniger model~\cite{Lieb_1963a, Lieb_1963b} --  a rare example of a solvable model describing a quantum many-body system -- plus its extension to nonzero temperature in the Yang-Yang model~\cite{YangYang1969}. These are used as the baseline for a number of successful \cite{Kheruntsyan03,Kheruntsyan05,Sykes08,Pietraszewicz_2017} and ongoing studies of fluctuations such as~\cite{Kerr2024}.  Still, fluctuations of $N_0$ and ensemble (in)equivalence were not subjects of research by this route.

The statistics of an interacting Bose gas has been the subject of many studies based on various Monte-Carlo methods~\cite{Krauth1996,Ceperley1999} most of which were focused on the interaction-induced shift to the critical temperature (see the review~\cite{Andersen2004}).
The full statistics of 1D trapped gas has been computed in~\cite{Carusotto2003} in the canonical ensemble confirming the results of the Bogoliubov approximation at low temperature.

\begin{figure}
    \centering
    \includegraphics[width=\columnwidth]{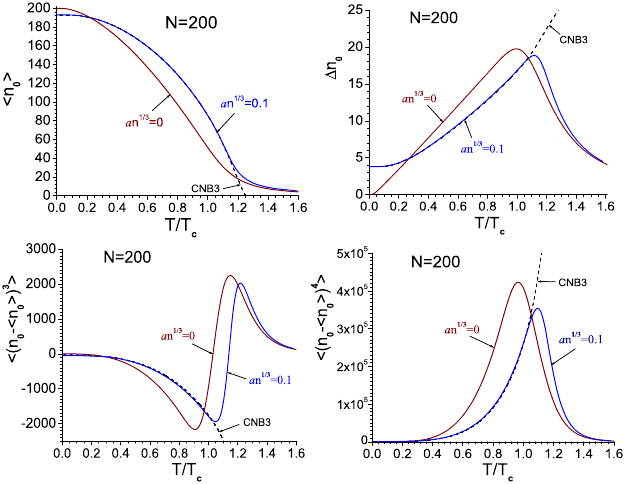}
    \caption{The first four moments of condensate occupation in a 3D box with $N=200$ in both the ideal gas (brown solid) and the weakly interacting gas ($an^{1/3}=0.1$, blue solid) as calculated using the hybrid master equation/quasiparticle approach~\cite{Svidzinsky2006}. The dashed ``CNB3'' lines refer to the interacting quasiparticle based approach in~\cite{Kocharovsky2000,Kocharovsky2000a}.  
    Reproduced with permission from~\cite{Svidzinsky2006}. \copyright American Physical Society. Used with permission.}
    \label{fig:higher-moments}
\end{figure}

Returning to explicit studies of $\delta N_0$, the work of~\cite{Svidzinsky2006} (and more rigorously~\cite{Svidzinsky2010}, but arguably less accurately~\cite{Dorfman11}) showed that one can hybridise the Bogoliubov approximation with the master equation method of~\cite{Scully99}, obtaining well-behaved properties also at temperatures of the order of, and across $T_c$. A complementary approach explicitly imposing CN restraints is presented in~\cite{Dorfman11}.
Fig.~\ref{fig:higher-moments} shows the first four moments of $N_0$ in a 3D box in both ideal and interacting gases, as calculated with this approach.

Other works investigated various aspects of the problem. Refs.~\cite{Tarasov20b,Tarasov18,Tarasov20,Tarasov21} looked at 3D traps and boxes and found condensate fluctuations to depend on boundary conditions in mesoscopic systems and even in the thermodynamic limit; Refs.~\cite{Tarasov20b,Tarasov19} made also forays towards stronger interactions.

Condensate fluctuations in the interacting gas have also been studied via a completely independent \emph{constructive} approach using a classical field representation of wavefunctions in the canonical ensemble,  first in~\cite{Witkowska2010Mar}, which provided a constructive Metropolis algorithm. Another classical field approach are the ergodic evolution methods described in Sec.~\ref{ERG}. In~\cite{Bienias11} a harmonically trapped 1D gas was studied, and it was found that the peak fluctuations grow and 
shift to lower values of $T/T_*$ as $g>0$ grows (along with an overall reduction of $\langle N_0\rangle$ for the same $T/T_*$, $T_*$ being the characteristic transition temperature for the system). 
The same effect was seen for a 3D trap by~\cite{Bhattacharyya2016} using a different approach with the help of recursion relations. However an opposite effect was reported for hard sphere pseudo-potentials~\cite{Hou20}.

Classical fields can also be easily applied to an interacting system with external particle exchange as in a grand canonical ensemble~\cite{Gardiner03,Proukakis08}, or intermediate ensembles~\cite{Pietraszewicz2017} that better approach experimental levels of fluctuations. The latter study constructively confirmed from an independent angle that the introduction of interactions suppresses the extravagantly large fluctuations seen in the ideal gas in a GCE. 

A separate topic of interest for which condensate fluctuations have been investigated despite the difficulties is the weakly \emph{attractively} interacting gas. First~\cite{Bienias11a} using a classical field approach for a 1D trapped system, found a small reduction of condensate fluctuations for weak attraction, turning into an increase as attraction became stronger. A later work used recursion relations~\cite{Bera17} on Li$^{7}$ parameters of trapped finite-size condensates, finding $\delta N_0$ fluctuations reduced compared to the non-interacting case.

One should remark about the restoration of ensemble equivalence by interactions. 
It is generally accepted that with sufficiently strong interactions the interactions energetically suppress any excessive number fluctuations~\cite{Wilkens97} and equivalence between GC and CN ensembles are restored. This can take place even for objectively very weak interactions if the system is collectively large.
The details of this matter have been, and continue to be, widely debated~\cite{Wilkens97,Giorgini1998,Idziaszek1999,Kocharovsky2000a,Yukalov2005a,Yukalov05b,Kocharovsky2006,Svidzinsky2006,Heller13,Cockburn11a,Touchette11,Squartini15,Tarasov15,Yukalov24,Baldovin18,Huveneers19,Kuwahara20}. 
The review~\cite{Kocharovsky2006}, chapter 6, gives some results regarding when ideal gas fluctuations transition to the interaction regime, using a Bogoliubov treatment.
Refs.~\cite{Kocharovsky16} and~\cite{Touchette15}  give fairly recent summaries of the current understanding, 
\cite{Yukalov24} is a recent view on the grand canonical catastrophe or, as named there, incorrect use of the GCE. 
The practical importance of the matter relates intimately to the question of whether one needs to be careful with one's choice of ensemble for theoretical or numerical calculations or whether it is fine to choose the most practically convenient one at one's disposal.

Finally, in real experiments with ultracold atoms it can occur also that variation in particle number due to technical elements of the preparation makes the total number fluctuations significant but less than what would be predicted by the GCE. For example,~\cite{JaskulaPhD} reported standard deviations $\delta N/N$ of about 20-35\%. 
To experimentally investigate the fluctuations of $N_0$ the deviation needs to be lower, for a typical total number of atoms, than $0.2$\%. The newer experiments with strong focus on control over $N$ report $0.01$\%~\cite{Kristensen2017} -- see Sec.~\ref{sec:experiments} for details.

Two further cases relevant with regard to intermediate ensembles 
are: (1) photonic BECs. Here the size of the particle reservoir with which they are in contact can be varied to experimentally study the crossover between the GCE and CE in a controlled way~\cite{Weiss16}.  
(2) the broad field of polaritonic condensates \cite{Klaas2018, Alnatah2024Mar} -- here the system is  very open in terms of particle exchange so that it can not be described as an equilibrium system even in its steady state.
Methods to explicitly deal with such intermediate ensembles include averaging of CN ensemble results with $N$ variation entered by hand, stochastic equations tailored to polariton physics \cite{Chiocchetta14,Bobrovska15} or automatic control of the $N$ variance while stochastically generating ensembles like in~\cite{Pietraszewicz2017}.

As we mentioned previously, the problem of fluctuations in the interacting Bose-gas is notoriously difficult. Most of the above-mentioned results were devoted to the canonical or grand canonical ensembles. 
On the other hand the recent experiment \cite{Christensen2021} reported fluctuations below the canonical one. 
In the next subsection we describe a recent advance -- a stochastic method suitable for weakly interacting systems in a microcanonical ensemble.

\subsection{The Hybrid Sampling Method}
\label{HYB}
Among the methods described in Sec.~\ref{sec:modern-frameworks} two in particular used a Metropolis sampling approach. In order to obtain the canonical ensemble in a wave picture (classical fields) or the complementary particle picture (Fock states sampling  - FSS).
Both 
scan the space of states to collect samples 
representative of the statistical ensemble from, 
a Markov chain constructed using the Boltzmann factor $e^{-\beta E}$ to determine the acceptance criterion in the Metropolis algorithm.  Here $E$ is the energy
of the non-interacting Hamiltonian. Both methods can perfectly reproduce exact results in the ideal gas canonical ensemble, with the added benefit in FSS method of calculating the condensate statistics in the microcanonical ensemble.

At first glance, including the effect of weak collisions in both methods seems simple:
ensure that the energy defining the Boltzmann factor includes the expectation value of the
interaction Hamiltonian. 
However, applying this approach to the most relevant
case of a harmonic traps yields different results for the classical fields and the FSS methods.
According to the classical fields method, collisions increase fluctuations, while the FSS
method indicates they decrease them.

In fact, both methods have serious drawbacks when applied to the interacting gas. The set of states in the FSS method contains only Fock states and lacks the vector space structure fundamental to coherent and entangled elements of quantum physics. There are no superpositions, no coherence. It works well for the ideal gas \cite{Kruk23,Christensen2021} in which a hard-wired basis set does not pose a problem, but there is no provision for interference or repulsion between locally overlapping orbitals if interaction is introduced.

The classical fields method operates in a vector space, dealing very well with amplitudes, superpositions, and not relying strongly on basis choice. The condensate wave function, as defined via Onsager-Penrose criterion \cite{Penrose1956}, can be computed post-factum by diagonalizing the one-particle density matrix, and subsequently the correct state occupation statistics can be calculated \cite{Brewczyk2007Jan}. However, the classical fields method does not deal well with occupations in the high energy modes which decay too slowly as $k_BT/\ve_j$ instead of the $1/(e^{\ve_j\beta}-1)$ required by full quantum mechanics in the particle-like regime.
It is therefore plagued by the ultraviolet catastrophe, with results heavily dependent on the cut-off value, whose best though not ideal value must vary nontrivially with temperature and the coupling constant \cite{Bienias11a,Blakie08,Pietraszewicz18b,Pietraszewicz18a}. The aim of this section is to present a hybrid sampling method, which combines the virtues of both methods: allowance of coherences and an adaptively shaped condensate wave function characteristic of the classical fields method, while preserving the absence of a necessary cut-off as inherited from the FSS method.

The set of states used to model the canonical ensemble consists of all classical fields in
consecutive single particle orbitals. In practical numerics, only a finite number of states is
necessary, so we use a sufficiently high energy cut-off. Thus, each state $\Psi(\bo{r})=\sum_{j=0}^{j_{\rm max}} \alpha_j \phi_j (\bm{r})$ is defined by a finite vector of
complex amplitudes
\begin{equation}
    \vec{\alpha}     = \bb{\alpha_0,\,\alpha_1,\ldots\alpha_{j\rm max}},
\end{equation}
just like in the classical fields approximation \eqref{cfPsi}, but now, the occupation
of each orbital is restricted to be an integer $N_j = |\alpha_j|^2$ with global constraint $\sum_j N_j = N $ -- this like in the FSS approach.

In the hybrid method, each proposed update 
in the Markov chain consists of two substeps. First, we
move one atom from orbital $j$ to orbital $l$ according to the FSS algorithm described
in Sec.~\ref{subsec:fssm}. Next, we rotate the phase of every nonzero (occupied) $\alpha_i$. To do so one first adds a random noise
\eq{hybrid_phase}{
    \alpha_i \mapsto \alpha_i + \delta\,\xi_i
}
where $\delta$ is a (optimally small, $|\delta| \ll 1$) parameter fixed mulation and $\xi_i$ a complex Gaussian random number of variance one, and then normalizes to the original $|\alpha_i|$ that one had after moving one atom $j\to l$. Overall this produces smaller relative phase changes in highly occupied orbitals.

 This way, we obtain a new state with a single atom transferred
from orbital $j$ to $l$ and with randomly modified   
phases.
As a next step, we calculate the
expectation value of the Hamiltonian in each state:
\begin{equation}
 H(\vec{\alpha}) = \sum_i E_i |\alpha_i|^2 + \frac{g}{2} \int d^D \bo{r} \left| \sum_{i} \alpha_i \phi_i(\bo{r}) \right| ^4
\end{equation}
where with the notation of \eqref{cfPsi},
\begin{equation}
    E_j = \int d^D\bo{r}\ \phi^*_j(\bo{r})\left[-\frac{1}{2}\nabla^2+V(\bo{r})\right]\phi_j(\bo{r}).
    \label{EU}
\end{equation}      
We chose $\phi_j(\mathbf{r})$ 
as the single particle orbitals of the empty trap, but choices may depend on the problem, as usual in classical field methods.

The Boltzmann factor $e^{-\beta H(\vec{\alpha})}$ 
is used in the usual Metropolis algorithm manner to compare to a random number $r\in[0,1)$ giving acceptance when $e^{-\beta H(\vec{\alpha})}\ge r$, cloning otherwise. 

Once the canonical ensemble is constructed from sufficiently spaced Markov chain samples $\vec{\alpha}^{(\nu)}$, $\nu=1,\dots,\mc{M}$, taking care to gather a large ensemble and discard the initial transient, the next step towards extracting the condensate \cite{Brewczyk2007Jan} is the construction of the one-particle density matrix, which can be done directly in the orbital basis
\begin{equation}
\rho_1(\vec{\alpha}, \vec{\alpha}^{\prime}) = \langle \alpha_j^{*} \alpha_{j\prime} \rangle =\frac{1}{\mc{M}}\sum_{\nu=1}^{\mc{M}} \alpha_j^{(\nu)*}\alpha_{j\prime}^{(\nu)}.  
\end{equation}

According to the Onsager-Penrose definition, the dominant eigenvalue $\lambda_0=N_0$ is the condensate fraction. The corresponding eigenvector $(\alpha_{0,0}, \alpha_{0, 1},\ldots, \alpha_{0, j_{\rm max}})$ determines the condensate wavefunction $\psi_{0}(\bm{r})=\sum_{j=0} \alpha_{0j} \phi_j(\bm{r})$.  
Then the probability
distribution of condensate occupation in the CN is represented by a histogram $\mathcal{P}(N_0^{(\nu)})$ of projections of the individual samples $\vec{\alpha}^{(\nu)}$ onto condensate: 
\eq{N0v}{
N_0^{(\nu)} = \sum_j \left|\alpha_j^{(\nu) *}\alpha_{0,j}\right|^2.
}
As explained in Sec.~\ref{subsec:fssm}, the postselection process on $H(\vec{\alpha}^{(\nu)})\in [\wb{E}-\Delta_E/2;\wb{E}+\Delta_E/2]$ allows us to also extract the
microcanonical statistics at energy $\wb{E}$.

\begin{figure}
    \centering
    \includegraphics[width=\linewidth]{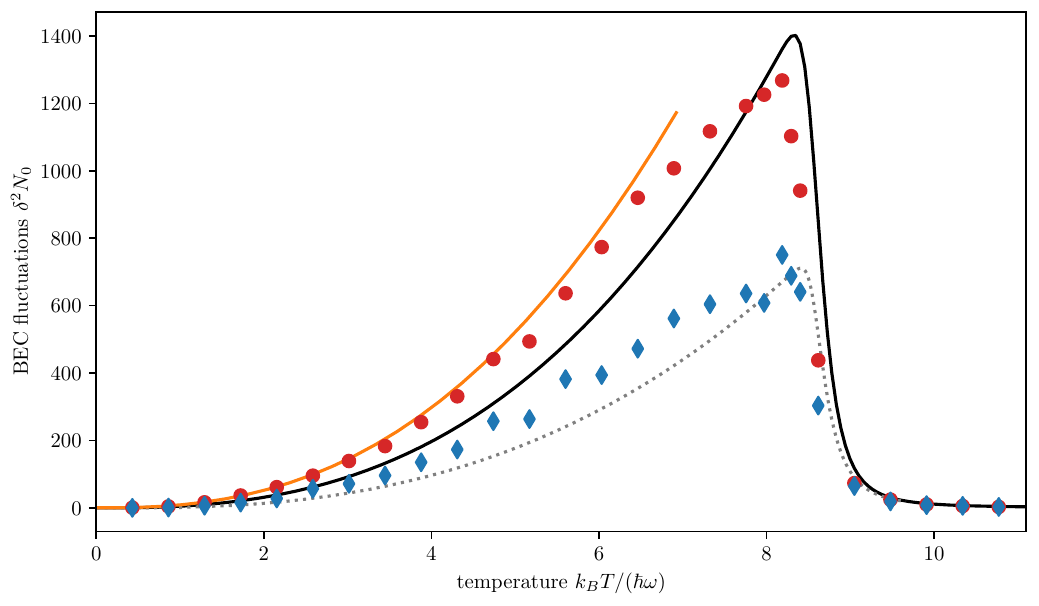}
    \caption{
    Fluctuations of the condensate atom number as a function of temperature in a 3D spherically harmonic trap with frequency $\omega=2\pi\times100$ Hz 
    for $N=1000$ Rubidium atoms (assumed scattering length $a=81.8$ Bohr radii)
    Solid black line -- exact canonical fluctuations for the noninteracting gas (\ref{eq:weiss-wilkens-recurrence}); dotted grey line -- exact micronanonical fluctuations for the noninteracting gas; solid orange line -- canonical fluctuations calculated with the Bogoliubov approximation; red circles -- hybrid method, canonical ensemble (note the near perfect agreement with Bogoliubov approximation at low temperatures); blue diamonds -- hybrid method, microcanonical ensemble.
\label{fig:hybrid_harmonic_3d}}
\end{figure}

We demonstrate the utility of the hybrid method with the physically relevant results shown in Fig.~\ref{fig:hybrid_harmonic_3d}. Specifically, we consider $N=1000$ rubidium atoms confined in a 3D spherically symmetric harmonic trap with a frequency of $\omega/2\pi=100$ Hz.
The data are represented as follows: black solid lines correspond to results for a non-interacting gas in the canonical ensemble, while black dotted lines show the microcanonical results for the same non-interacting gas. Both of these results are exact.
The results for the interacting gas were obtained using the hybrid method. Red circles represent canonical data, while blue diamonds denote microcanonical data. The orange dotted lines correspond to the benchmark of the (particle non-conserving) Bogoliubov approximation. 
The hybrid CN and Bogoliubov results align at low temperatures. We see that in this case, interactions reduce the maximum fluctuations observed in the canonical ensemble.
Notably, while interacting CN results in this regime with large $N$ and $T$ might also be obtained using a master equation approach, this is the only current approach that gives MC fluctuations in this regime (barring a basic classical field ergodic method which would have  large cutoff-dependent systematics in 3d).

The methods just discussed 
have been used for some comparisons with experimental results, but their number remains limited and the comparisons have been quite noisy. One significant reason for this is the scarcity of experimental data: to date, only one group has achieved the necessary precision in measuring fluctuations in the number of condensed atoms. Additionally, the large number of atoms typically involved in such experiments poses significant challenges for the current theoretical/numerical methods.

\section{Experiments\label{sec:experiments}}

In the following, we review the experimental progress in addressing the measurement of condensate fluctuations.

The experimental realization of weakly interacting BECs introduced a new framework for studying many-particle quantum systems. However, until recently the experimental investigation of BEC population statistics was limited to the first moment, and higher moments, such as BEC atom number fluctuations, remained inaccessible due to technical noise. The considerable theoretical interest in number fluctuations discussed in this review crucially necessitates the experimental exploration of this fundamental feature.

Two recent experimental developments have allowed for a deeper investigation of the fluctuations in Bose gases. On the one hand, atomic BECs have made considerable progress in stability by using active monitoring and feedback techniques. This has allowed for a suppression of the technical noise contributions, revealing the inherent microcanonical fluctuations~\cite{Christensen2021}. On the other hand, the creation of photonic BECs~\cite{klaersBoseEinsteinCondensation2010a} has provided access to non-interacting Bose gases which realize the grand canonical scenario. We will emphasize the former case since recent theoretical progress has focused on the interacting Bose gas.

\subsection{Experiments with atomic Bose gases}

\subsubsection{Relevant exprimental developments}

Until recently, primarily the number of particles in a BEC, corresponding to the first moment of the statistical distribution had been investigated~\cite{Gerbier2004,Meppelink2010,Tammuz2011}. This provided access to the fraction of condensed atoms as a function of temperature and thus allowed for evaluation of the critical temperature~\cite{Gerbier2004a,Smith2011}. Typically, the ideal gas model already captures its general behavior in experiments with ultracold atoms. The agreement is improved by the inclusion of corrections for interactions and finite size effects~\cite{Dalfovo1999}. Importantly, interactions induce a shift of the critical temperature, which can be understood intuitively; A positive scattering length causes the atoms to repel each other, reducing the central density in a ultracold cloud, which in turn reduces the critical temperature for condensation. Note however, that higher order corrections push the shift in the opposite direction~\cite{Davis2006}. A more detailed comparison between experiment and theory can be achieved by treating the interactions between the condensate atoms and the thermal atoms with self-consistent methods. Detailed experimental investigations of the condensate fraction were reported in~\cite{Gerbier2004,Meppelink2010,Tammuz2011,Smith2011}. The observation of beyond mean-field contributions to both the shift in the critical temperature and the condensate fractions was reported in~\cite{Gerbier2004a,Smith2011b}. A wide array of experiments have also probed local density fluctuations, some of the first being \cite{Esteve2006,Jacqmin11}. However, local fluctuations do not provide a way to extract the fluctuations of the number of atoms in the ground state mode -- which is nonlocal.

\begin{figure}
    \centering
    \includegraphics[width=\linewidth]{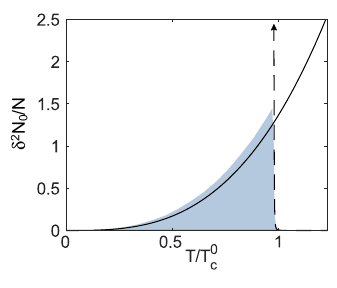}
    \caption{Limiting cases of the expected fluctuations in experimental realizations, for the case of an ideal gas with $N\sim5\times10^5$ in an isotropic harmonic trap. At the critical temperature $T_\mathrm{c}^0$ a grand canonical ensemble calculation predicts unphysically large fluctuations (dashed line). Below $T_\mathrm{c}^0$ the leading CN contribution of Eq.~\eqref{CN_fluc} provides a good estimate of the fluctuations~\cite{Politzer1996,Navez1997} (solid line). An exact numerical calculation~\cite{Weiss1997} shows how these two expressions serve as limiting cases in the two temperature regions~(shaded area). Reproduced from~\cite{Kristensen2019}
    \copyright American Physical Society.
    Used with permission.
    \label{fig:LimitingCases}}   
\end{figure}

The first time the statistical properties of a Bose gas were considered experimentally was a seminal experiment showing the sub-Poissonian number statistics in such a gas~\cite{Chuu05}. These measurements were conducted for very low total atom numbers $50<N<3000$ at low temperatures. In the regime $N<400$ they showed sub-Poissonian statistics with $\fluct{N}/N<1$ as expected from analogy to Fig.~\ref{fig:LimitingCases}. However, the experiment did not distinguish between condensed and thermal atoms and thus it remained unclear whether the effects of the evaporation process or the inherent fluctuations of the condensed atom number had been observed. 

Importantly, fluctuations in BECs are also caused by nonequilibrium phenomena that go beyond standard statistical ensembles. 
In particular, phase fluctuations typically occur in BECs at high temperature, especially in elongated or low-dimensional clouds \cite{Petrov00,Petrov01}, and lead to density modulations after time-of-flight (TOF) imaging~\cite{Dettmer2001}. Furthermore, studies have investigated the dynamical behaviour of the condensate fraction when crossing the phase transition. Wolswijk~{\sl et al.}~\cite{Wolswijk22} measured fluctuations of condensate fraction in an Na$^{23}$ during evaporative cooling as it took the gas below $T_c$. They found exponential decay of the fluctuations at a rate somewhat slower than the condensate growth rate. 

\subsubsection{Challenges}

The challenges for the experimental observation of fluctuations in a typical quantum gas experiment are 
directly evident from fluctuation calculations illustrated by Fig.~\ref{fig:LimitingCases}. Consider a sample with total atom number $N\sim 5\times10^5$. If the peak fluctuations are of the order of $\sqrt{N}$, this means that the standard deviation of the BEC atom number $ N_0 $ has to be measured at a level of $\delta N_0\approx 700$. However, any drift in the temperature $ T $ and atom number $ N $ also leads to a change in $N_0$. The stability requirements for $ N $ and $ T $ can be estimated from the well-known ideal-gas result Eq.~\eqref{cond_frac_har} for the condensate atom number, $ N_0 = N - {\rm const.}\times T^3 $. 
To first order the variation in $ N_0 $ for low condensate fraction is then $ \delta N_0 \sim N(\frac{\delta N}{N} - 3\frac{\delta T}{T})$. The relative stability of $ N $ and $ T $ must thus meet the condition $ \frac{\delta N}{N} < \SI{0.14}{\percent} $ and $ \frac{\delta T}{T} < \SI{0.05}{\percent} $. In principle, this can be achieved in two ways. Either the experiment inherently meets the stability criteria, or alternatively, it may be possible to post select data within a small range of $ N $ and $ T $ in an experiment with large shot-to-shot fluctuations but very good imaging allowing for precise determination of $N$, $N_0$ and $T$. Recent work has shown that only the first option is possible~\cite{Vibel2024}, since post-selection adds an ambiguity to the measured ensemble precluding the identification of fluctuations. Additionally, the post-selection method would require a very large number of realizations due to the variation of $N$ and $N_0$ by $10\%$ in typical experiments. Therefore, the active stabilization of the experiment was chosen in recent experimental realizations outlined below.

In addition to preparation, experimental noise is present in the measurement and estimation of the atom numbers. Noise sources include, but are not limited to, shot noise in imaging light, camera noise, optical imperfections, fitting uncertainties, light frequency variations, the probabilistic distribution of a finite number of atoms, and the effects of interactions between atoms on the distributions used to extract the thermal and BEC atom numbers individually. Moreover, these noise sources have a complex dependency on both atom numbers and temperature. These effects have to be taken into account to an appropriate extent when the atom number fluctuations are measured~\cite{Vibel2024}.

\subsubsection{Experimental Improvements}

To improve the preparation noise in the experiment with active stabilization it is necessary to extract information about the system non-destructively. The dark-field Faraday imaging (DFFI) technique reported in~\cite{Gajdacz2013} is well suited for this task. The technique exploits the Faraday rotation effect of the polarization of light traversing an atomic sample. By using far-detuned probe light, the absorption of light is limited and the atomic sample is left relatively unaltered. In~\cite{Gajdacz2016}, the probe light was linearly polarized and the non-rotated light was diverted towards a beam dump by a polarizer. Consequently, only the rotated probe light passes the polarizer and was detected on a CCD camera. This then allowed for spatially resolved non-destructive images of ultracold clouds and in particular for the extraction of the atom number. This dark-field Faraday imaging approach was also shown to yield a precision that surpasses the atom-shot-noise~\cite{Kristensen2017} for appropriate imaging conditions.

By adding an active feedback mechanism to the DFFI technique, the atom number in a single shot can be decreased in a controlled manner and thus the atom numbers can be stabilized at the cost of slightly fewer atoms. This method was demonstrated in~\cite{Gajdacz2016} where the radio frequency (rf) evaporation in a magnetic trap was controlled in real-time in response the DFFI signal. The stabilization method was later implemented in~\cite{Kristensen2019} to achieve the first measurements on the atom number fluctuations in partly condensed BECs.

\subsubsection{Experimental realization in ultracold Bose gases}

In the following, we briefly outline the experimental steps used in the recent observations of fluctuations in BECs~\cite{Kristensen2019,Christensen2021,Vibel2024}. To initiate the experiments $\approx 10^9$ $^{87}\text{Rb}$ atoms were captured and cooled in a magneto-optical trap. The cloud was then optically pumped into the $\ket{F = 2,m_F = 2}$ state and transferred to a harmonic magnetic trap with axial frequency $\omega_a$ and radial frequency $\omega_\rho$. In this trap, further rf evaporative cooling was performed to lower the temperature of the sample. This sequence is prone to technical fluctuations that can obscure the observation of atom number fluctuations. To address this, the stabilization procedure outlined above was initiated when the cloud contained $\sim 4 \times 10^6$ atoms at a temperature of 14~$\mu4$K. The cloud was probed using minimally destructive Faraday imaging~\cite{Gajdacz2013,Kristensen2017} to determine the atom number in that particular experimental realization. This number was then corrected with a weak rf pulse of controlled duration which removed excess atoms. Subsequently, a second Faraday measurement verified the outcome, thereby achieving a relative stability of atom number at the $10^{-4}$ level~\cite{Gajdacz2016}. 
For the final cooling step towards BEC, a tightly compressed magnetic trap with a high collision rate was used. In its most compressed state, the trap's aspect ratio was $\lambda = 174$, which can cause significant phase fluctuations across the cloud that evolve into density modulations during time of flight, complicating precise atom number and temperature determination~\cite{Dettmer2001}.
To mitigate this, the trap was decompressed for measurements, reducing the aspect ratio to values between $4.5 < \lambda < 10$, where the phase coherence length exceeds the condensate length along the long axis. In the final stage, BECs were produced by forced evaporation at a radio frequency corresponding to the desired BEC temperature. To ensure thermal equilibrium, the BEC was held at the final frequency for 800~ms, followed by another 400~ms without rf radiation, before the trap was switched off. The cloud was then probed after a time-of-flight expansion using resonant absorption imaging on the $\ket{F = 2} \rightarrow \ket{F' = 3}$ cycling transition. A measurement consisted of 4 images used to reduce undesired imaging artifacts. Image 1 contained atoms, image 2 had no atoms, and image 3 and 4 had no imaging light but were otherwise taken under the same conditions as image 1 and 2. Image 3 was subtracted from image 1 and image 4 from image 2 to remove dark counts from the camera. 

Great care was taken to mitigate imaging effects on the detected atom number. This includes calibration of the effective saturation intensity~\cite{Reinaudi2007,Hueck2017,Veyron2022,Vibel2024alpha}, the radius of the thermal cloud fitting region~\cite{Szczepkowski2009}, and the effects of the imaging photons on the cloud~\cite{Ketterle1999,Muessel2013}. Effects of imaging induced Doppler shifts were also considered~\cite{Genkina2015,Hueck2017}. The focusing of the imaging system was optimized by following previously developed methods~\cite{Putra2014}. The imaging light parameters were optimized to minimize noise according to earlier work~\cite{Pappa2011,Horikoshi2017,Genkina2015,Lueckes_PhD}. Techniques to reduce shot noise and imaging fringes were considered~\cite{Ockeloen2010,Niu2018, Ness2020, Song2020}. It proved especially important to keep the imaging duration short to avoid any movement during the image. Keeping the duration between images short reduces the fringes caused by oscillations in optical elements when comparing the images with and without atoms.

To determine the BEC atom number in each image, the wings of the cloud are fitted with a Bose-enhanced thermal distribution reduced to the two-dimensional column density from which the temperature is obtained. The fitted distribution is subtracted from the image, and the BEC atom number $N_0$ is obtained by integration of the remaining column density. 

However, the variance cannot be determined directly from $N_0$, since small remaining drifts of the magnetic offset field lead to minor temperature variations with a median standard deviation of $\sim \SI{3}{\nano\kelvin}$. This drift is eliminated by subtracting a linear fit of the condensate number as a function of the total atom number~\cite{Kristensen2019} and determine the residuals $\eta_i$, where $i$ indicates the order in time.
The BEC atom number variance is then given by a two-sample variance of the residuals
\begin{equation}
\label{eq:TwoSampleVariance}
	\fluct{N_0} = \frac{1}{2}\left\langle \left(\eta_{i+1} - \eta_i\right)^2 \right\rangle.
\end{equation}
Thus, this two-sample variance contains the BEC fluctuations and detection noise, but excludes slow technical drifts. This evaluation technique is commonly called the \textit{correlation method}.

The peak fluctuations are determined by a fit to $\fluct{N_0}$ based on the theoretical expectation illustrated in Figs.~\ref{fig:graphical-abstract}~(b), \ref{fig:Holthausprl}, and~\ref{fig:LimitingCases}. Specifically the fit function mimics the asymptotic behavior of the fluctuations in a non-interacting gas $ \fluct{N_0} = \zeta(2) \left(\frac{k_\mathrm{B}}{\hbar\overline{\omega}}\right)^3T^3$, where $\overline{\omega}$ is the geometric mean of the trapping frequencies~\cite{Politzer1996}. Moreover, the fluctuations decay in near step-like manner close to the critical temperature, which we model with a Heaviside step function $\Theta(T_p - T)$, where $T_\mathrm{p}$ is the temperature at the peak fluctuations. 
To account for small temperature drifts the expression is furthermore convolved with a normal distribution $\mathcal{N}(T,\sigma_T)$ centered on the temperature $T$ with a standard deviation $\sigma_T$ given by the median of the measured temperature variation. Thus, fit model is giveb by
	\begin{equation}\label{eq:fitModel}
		\fluct{N_0}(T) = \left(f\ast g\right)(T) +\mathcal{O}
	\end{equation}
	where $f$ and $g$ are given by
        \begin{eqnarray*}
		f(T) &= \delta N_\mathrm{0,p}^2 \left(\frac{T}{T_\mathrm{p}}\right)^3 \Theta(T_\mathrm{p}-T),
		\\
		g(T) &= \mathcal{N}(T,\sigma_T).
        \end{eqnarray*}
The three fit parameters are the peak atom number variance $\delta N_{\mathrm{0,p}}^2$, $T_\mathrm{p}$ and an offset $\mathcal{O}$ which accounts for experimental noise~\cite{Kristensen2019}. They constitute the final result of an experiment at a given initial atom number and trap configuration. A more refined fit has recently been explored in~\cite{Vibel2024}.

\subsubsection{Experimental results with ultracold gases}
The first observation of atom number fluctuation in a BEC according to the methods outlined above succeeded in 2019~\cite{Kristensen2019}. The stabilization technique based on Faraday imaging enabled the preparation of ultracold thermal clouds at the shot noise level, eliminating numerous technical noise sources. In the data analysis, the \textit{correlation method} was utilized to accurately determine both the fluctuations and the sample temperature.

\begin{figure}
    \centering
    \includegraphics[width=\linewidth]{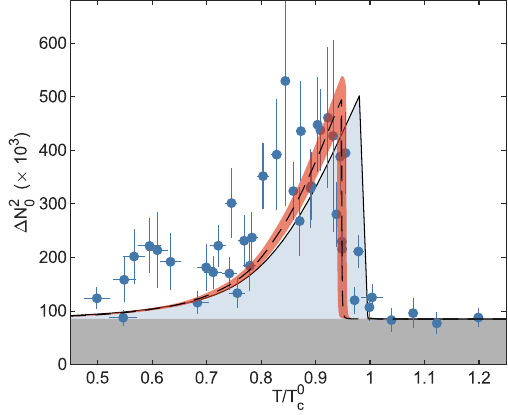}
    \caption{Variance of the BEC atom number, constituting the first detection of BEC atom number fluctuations. The dashed line is a fit to the data (see text) where shading indicates the confidence bound on the fit. The data is compared to a prediction in a non-interacting gas (light blue shading). Gray shading indicates a fitted offset due to technical fluctuations. 
    Reproduced from~\cite{Kristensen2019}
    \copyright American Physical Society.
    Used with permission.
    }
    \label{fig:Fluctuations}
\end{figure}

This allowed for the observation of the distinctive signature of the fluctuations, namely a sudden increase in fluctuations of the condensate atom number near the critical temperature as shown in Fig.~\ref{fig:Fluctuations}. Note that these initial experiments employed a less refined fit function than outlined above. Nonetheless, they captured both the size of the peak fluctuations and the shift of the critical temperature due to interactions in the system. 

\begin{figure}
    \centering
    \includegraphics[width=\linewidth]{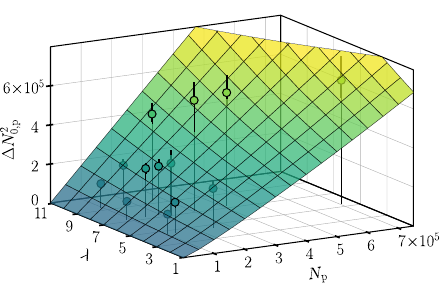}
    \caption{Peak variance of the BEC atom number as a function of $N_\mathrm{p}$ and trap aspect ratio $\lambda$. The experimental data was compared to a theoretical expectation for the non-interaction canonical ensemble, scaled to the data in a fitting procedure (see text).
    Reproduced from~\cite{Christensen2021}. \href{https://creativecommons.org/licenses/by/4.0/}{CC BY 4.0}
    }
    \label{fig:ExpMicro}
\end{figure}

These initial experiments raised the question whether the effects of interaction and the appropriate thermodynamic ensemble can be identified. This was investigated by measuring the peak fluctuations for various atom numbers and trap geometries~\cite{Christensen2021} as shown in Fig.~\ref{fig:ExpMicro}. Most significantly, the study showed a reduction in observed peak atom number fluctuations by 27\% with respect to predictions for a canonical non-interacting gas. This reduction is expected in the microcanonical case and thus underscores the microcanonical nature of the system. The experiment also showed anomalous scaling of fluctuations with atom number, which was attributed to the interplay between dimensional and interaction effects. These results were accompanied by simulations showing the microcaonical reduction of fluctuations and the dependence on trapping geometry and atom number. This work advanced the field by demonstrating the critical role of the microcanonical ensemble in describing fluctuations and thus highlighted the path towards the interpretation of future experiments with ultracold gases.

\begin{figure}
    \centering
    \includegraphics[width=\linewidth]{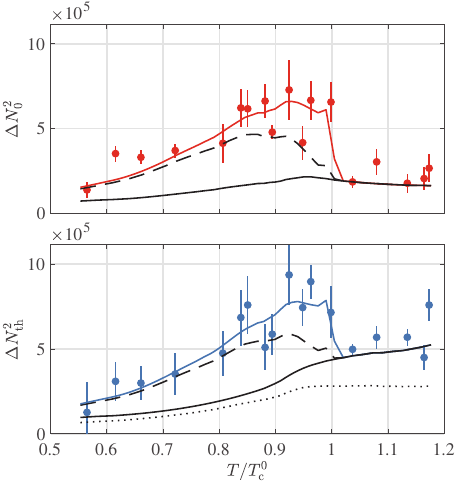}
    \caption{Fluctuations in BECs and the corresponding thermal clouds. Upper panel: Variance of the BEC atom number as a function of temperature. The variance due to estimation errors (full black line), and total technical noise (dashed black line) are indicated. The colored line shows the total fitted variance. Lower panel: Variance of the thermal atom number shown the with the same contributions to the variance. In addition, a dotted line indicates the estimation error extracted from the Monte Carlo simulation. 
     Reproduced from~\cite{Vibel2024}. \href{https://creativecommons.org/licenses/by/4.0/}{CC BY 4.0}.
    }
    \label{fig:dualfluct}
\end{figure}

The most recent work on fluctuations in degenerate Bose gases analysed the correlations between the thermal and the BEC atom number~\cite{Vibel2024} in a partially condensed Bose gas. If only fundamental fluctuations were present, one would naturally expect a perfect anticorrelation. The possible observation of such a correlation thus served as a motivation for the investigations. 

To allow for a comparison of the variances in the thermal and the condensed could, a comprehensive model of the noise sources was developed. This allowed for a distinction between noise contributions due the preparation, the image analysis and the inherent fluctuations. The three largest noise contributions in the analysis procedure were identified as the effect of the image fitting technique, the effect of the binomial distribution of atoms in each pixel, and the effect of the variation of the imaging laser frequency. These uncertainties, together with several smaller contributions, were included in the analysis of atom number fluctuation results for the thermal and BEC components as shown in Fig.~\ref{fig:dualfluct}. This led to a temperature-dependent offset of the variance due to all technical noise sources and complicated the analysis, in particular since the technical noise features a slow increase starting at the critical temperature. In total the analysis provided an improved value of the peak BEC atom number fluctuations, $\delta N_\mathrm{p,0}^2 = (3.7\pm7)\times10^5$ near the critical temperature for the specific experimental parameters, representing a 41\% reduction compared to Ref.~\cite{Christensen2021} and further supports the microcanonical nature of the fluctuations.

The noise analysis also revealed that the \textit{correlation method} to extract the variance leads to an artificial anticorrelation between the thermal and BEC atom numbers. Hence, the measurement of atom number fluctuations in the thermal cloud as shown in Fig.~\ref{fig:dualfluct} can not be considered an independent observation of fluctuations but presents a complementary method to obtain fluctuation data. 

These recent achievements highlight the potential for more refined experiments, including larger atomic samples to approach the thermodynamic limit, and improved measurement techniques to reduce uncertainties. This research provides a foundation for an improved understanding the complex interplay of quantum statistics, interactions, and ensemble constraints in Bose gases. 

\subsection{Experiments with Photons}

Photonic Bose-Einstein condensates in dye microcavities~\cite{klaersBoseEinsteinCondensation2010a} provide access to the investigation of condensate statistics in different ensembles from atomic condensates. The photonic condensates can have access to controlled heat and particle reservoirs in the dye molecules resulting in in systems described by describes canonical or grand canonical ensembles. In particular photonic condensates are often compared to lasers, but the two systems exhibit some fundamentally different properties with fluctuations being one of them. 

Early theoretical investigations of photonic condensates in the grand canonical ensemble showed fluctuations above Poisson-like behaviour and increased fluctuations compared to atomic experiments~\cite{Klaers2012}. Shortly after, fluctuations in photonic condensates were observed experimentally~\cite{Schmitt14}. The experiment yielded both the fluctuations and the probability distributions of photonic condensates for different condensate fractions. The observations found the expected unusually large fluctuations and a comparison with the non-interacting grand canonical predictions found a good agreement in the measured range.

Theoretical calculations of the effects of contact interactions in photonic condensates were carried out~\cite{Wurff14} and compared to the experimental results from~\cite{Schmitt14}. The result was a good agreement with the experiment but stressed the importance of a systematic, experimental investigation of the interaction strength. Further theoretical investigation of the effect of weak interactions in experimentally accessible few-particle systems were carried out with two different models in~\cite{Weiss16}. The resulting fluctuations for grand canonical ensembles were still much higher than for canonical ensembles. It was also found that the typical textbook treatment using the ideal gas model differed significantly from even weakly interacting systems. 

The fluctuation-dissipation theorem which connects the thermal fluctuations in a system to its response to external perturbations were experimentally investigated in~\cite{Ozturk23} using photonic condensates. The response and the fluctuations of the photonic condensates were measured separately and compared. The result was good agreement with the theorem and demonstrated the thermal nature of the photonic condensate system. 

The dynamics of fluctuations in open photonic condensate systems were theoretically and experimentally investigated in~\cite{Ozturk19}. It was found that the precise behaviour of the fluctuations depended sensitively on the openness of the system and further study of the effect of interactions between system and environment was encouraged.

The experimental results with photonic condensates in~\cite{Schmitt14, Ozturk23} demonstrate their versatility in statistical remeasurements and the complementarity to atomic condensate experiments in investigation of different thermodynamical regimes. 




\section{Perspectives}
\label{sec:perspectives}

This review provides a comprehensive overview of the current theoretical understanding and experimental results on the condensate fluctuations in quantum degenerate Bose gases. We highlighted the available theoretical approaches with a special focus on the role of the ensemble choice. We discussed the important experimental progress made recently using two experimental platforms based on ultracold atom and photons in dye filed microcavities. Thus this work also updates previous reviews covering related topics~\cite{Ziff1977,Kocharovsky2006,Yukalov04,Yukalov24}. 

As we have seen in previous sections, numerous questions remain, providing opportunities for further investigation. In the following we first outline the perspectives for future experimental work and then discuss the next steps to be taken in theoretical development.

Based on the available techniques of producing highly stable Bose-Einstein condensates it should be possible to examine the higher moments of the fluctuations as shown theoretically in Fig.~\ref{fig:higher-moments}. However, the measurement of higher moments requires considerably more data and consequently a longer period in which the atom number and trap parameters are kept stable. Another avenue based on the existing experimental methods is the investigation of lower dimensional systems such as quasi-1D or quasi-2D systems where BECs can be produced with standard techniques. If the stabilization technique is combined with the use of Feshbach resonances to tune the interaction strength a realization close to the case of the ideal gas may even be possible. 

Another fascinating possibility is to explore fluctuations in experiments which can count individual atoms. In such experiments with small BECs it may be possible to freeze the entire ensemble and record each atom's position. Steps towards this goal have been taken in 2D systems recently~\cite{Xiang2024Nov,Verstraten2024Apr,yao2024} yielding the promise that exact counting experiments of fluctuations are on the horizon.

Moreover, the investigation of fluctuations can be extended to the growing number of available ultracold quantum systems. This includes multi component systems consisting of different atomic species or atoms in a number of quantum states~\cite{Baroni_2024}. In such quantum mixtures of ultracold gases the situation becomes yet more complicated due to the mutual interaction of the components. 
Alternatively, other interactions within a single component have become experimentally accessible, particular long range dipole-dipole interactions~\cite{Chomaz_2023}. These systems are currently under intense experimental scrutiny, and they would be a prime candidate to investigate the effect of such long range interaction on the fluctuations in a quantum gas. This study can be done together by both experiment and theory with established methods for
weakly interacting gases.

On the theoretical side, a similarly large number of questions remain unanswered. One direction of investigation is ensemble (in)equivalence. As discussed in Secs. \ref{sec:fluct-ideal-gas} and \ref{sec:modern-frameworks}, different ensembles lead to different results when it comes to condensate fluctuations in the ideal gas, even in the thermodynamic limit. This may change drastically for interacting systems. Interactions make density fluctuations energetically unfavourable, and it is expected that condensate fluctuations will become ensemble-independent. A definitive understanding of the anomalous fluctuation controversy in weakly interacting gases, such as the conditions under which the various results become valid as the thermodynamic limit is approached, has also not been fully reached. 

The Lieb-Liniger (LL) model \cite{Lieb_1963a, Lieb_1963b,YangYang1969}, a solvable model of a 1D homogeneous gas of atoms interacting via a delta potential, provides a natural system for obtaining rigorous results in this context. While fluctuation results exist for the LL gas at finite temperature \cite{Panfil2014,Kerr2024,Pietraszewicz_2017}, they neither address $N_0$ fluctuations nor differences between statistical ensembles. Even within a given ensemble, the role of interactions in anomalous fluctuations remains controversial and requires unambiguous resolution.
Another promising avenue for widening the topic lies in studying quantum statistics across different physical systems, including multicomponent condensates. Dipolar gases, known for exhibiting quantum droplets \cite{barbut2016, chomaz2016} and supersolids \cite{Guo2019Oct, Chomaz2019} have emerged as a highly active area of research in ultracold atoms. 
While some results exist on thermal effects in dipolar 
Bose-Einstein condensates (BECs), including changes in the BEC fraction at finite temperature \cite{Ronen2007, Ticknor2012, Bisset2012, Pawlowski2013, Boudjemaa2015Jan, Aybar2019Jan}, the fluctuations of the number of condensed atoms have not yet been investigated. Of particular interest are the gas-droplet transition and the self-cooling mechanism observed in quantum droplets \cite{Petrov2015}. Beyond the thermodynamic equilibrium state, significant attention has been devoted to atoms in resonant cavities and interacting via photon-mediated interactions \cite{Landig16,Brennecke2007,Baumann10},
or polaritonic BECs\cite{Kasprzak06,Balili07,Wertz09},
where dynamical equilibrium can lead to novel phenomena. 

Statistical properties of ultracold gases depend strongly on the experimental protocols of cooling and thermalization \cite{Christensen2021} and ultimately, none of the typically used paradigmatic ensembles fits the reality perfectly. Therefore, some intermediate ensembles should be used, either by relaxing the constraint on the extensive variables \cite{Pietraszewicz2017}, post-selection, or studying dynamics of an open system mimicking cooling processes \cite{Cirac1994}. Apart from this experiment matching need, there appear to be fundamental theoretical questions related to the system in equilibrium, but subject to other constraints. Here, the notion of the Gibbs ensemble has been used, with great success overall in recent years \cite{Bouchoule_2022}.

The research areas outlined above require the development of more sophisticated numerical methods. Even for the ideal gas, the calculation of statistics for experimentally relevant systems containing several hundred thousand atoms remains computationally challenging. Importantly, as shown in Figure \ref{fig:KR7} the thermodynamic limit is not achieved even for such large systems. Advances in methods for open quantum systems, generalizations of the Fock state sampling method of (FSSM), and new techniques within the framework of generalized hydrodynamics are expected to have a significant impact. 

Finally, as experimental precision increases, it becomes crucial to consider the practical consequences of ensemble dependence. Experimental observations indicate a reduction in BEC fluctuations by 27\% compared to canonical predictions. The question arises: do other quantities, such as magetization (fraction of subcomponents in a multicomponent condensate), phase correlations, density-density correlations, or recombination rates, also exhibit ensemble dependence in finite systems? Low-temperature results \cite{Sinatra2007} demonstrate that the rate of phase collapse can exhibit different scalings with the total number of atoms across different statistical ensembles. This finding highlights the importance of selecting the correct statistical ensemble in applications such as interferometry to ensure trustworthy results. 

The experimental and theoretical findings available to date have made significant contributions to the understanding of quantum statistical mechanics. The reduction in fluctuations and the observed scaling behavior offer critical insights into the unique properties of isolated quantum systems. 

This research also sets benchmarks for future studies, particularly in extending the analysis to higher moments of the fluctuation distribution and exploring time-dependent dynamics. Moreover, the theoretical methods and results will guide further investigations into interaction effects and their role in fluctuation phenomena.

\section{Acknowledgments}

We 
acknowledge support from the (Polish) National Science Center
Grants No. 2021/43/B/ST2/01426 (K.~R. and P.~K.),  2018/31/B/ST2/01871 (P.~D.), 2022/45/N/ST2/03511 (M.~B.~K.)..
M. F. A. and J. J. A. acknowledge support from the Danish National Research Foundation through the Center of Excellence “CCQ” (DNRF152) and by the Novo Nordisk Foundation NERD grant (Grantno. NNF22OC0075986).
K.P. acknowledges support
from the (Polish) National Science Center Grant No.
2024/53/B/ST2/02161.

\section*{Bibliography}

\end{document}